\documentclass{article}
\usepackage{naturetex}
\usepackage{subfiles} % Best loaded last in the preamble

\usepackage{hyperref}
\usepackage{nameref,zref-xr}
\zxrsetup{toltxlabel}
\zexternaldocument*{sup}

\begin{document}

% % COVER LETTER
% \subfile{cover_letter}
% \newpage

\maketitle

%\linenumbers

{\setstretch{1.0}
	% *** ABSTRACT ***
	\section*{Abstract}
The Oort cloud is thought to be a reservoir of icy planetesimals and the source of long-period comets (LPCs) implanted from the outer Solar System during the time of giant planet formation. The abundance of rocky ice-free bodies is a key diagnostic of Solar System formation models as it can distinguish between ``massive" and ``depleted" proto-asteroid belt scenarios and thus disentangle competing planet formation models. Here we report a direct observation of a decimeter-sized ($\sim2$ kg) rocky meteoroid on a retrograde LPC orbit ($e \approx 1.0$, i = $121^{\circ}$). During its flight, it fragmented at dynamic pressures similar to fireballs dropping ordinary chondrite meteorites. A numerical ablation model fit produces bulk density and ablation properties also consistent with asteroidal meteoroids. We estimate the flux of rocky objects impacting Earth from the Oort cloud to be $1.08^{+2.81}_{-0.95}~\mathrm{meteoroids/10^6~km^2/yr}$ to a mass limit of 10~g. This corresponds to an abundance of rocky meteoroids of $\sim6^{+13}_{-5}$\% of all objects originating in the Oort cloud and impacting Earth to these masses. Our result gives support to migration-based dynamical models of the formation of the Solar System which predict that significant rocky material is implanted in the Oort cloud, a result not explained by traditional Solar System formation models.
}% setstretch

% *** Main body ***
\newpage

% Nature Astronomy submission types:
% - Letters - max 2000 words, max 4 figures, max 30 references, 200 word summary
% - Article - 3k-3.5k words, max 6-8 figures, max 50 references
%  - unreferenced abstract, 150 words max
%  - An introduction (without heading) of up to 500 words of referenced text expands on the background to the work (some overlap with the summary is acceptable)
%  - concise, focused account of the findings (headed 'Results')
%  - one or two short paragraphs of discussion (headed 'Discussion')
% + Methods section which doesn't have any limits

\section*{Introduction}

% - An introduction (without heading) of up to 500 words of referenced text expands on the background to the work (some overlap with the summary is acceptable)

% Per paragraph (each 100-150 words):
% - general introduction - set the stage and present the main problem ("However, ...")
% - more details and background (e.g. theoretical models of solar system formation, material distribution, some observations)
% - more details and background (e.g. general distribution of material, previous surveys of strengths)
% - introduce observations and what was measured

The sharp increase in the number of ground-based networks utilizing digital cameras for observing fireballs \cite{spurny2017discovery, toth2019amos, devillepoix2020global, colas2020fripon, vida2021global} in recent years has resulted in near continuous coverage of almost $2\%$ of the Earth's atmosphere for small impactors. Supplementing these ground-based instruments in fireball detection is the Geostationary Lightning Mapper (GLM) instrument on board the GOES-16 and 17 satellites. First deployed in 2016, GLM now observes a total of $\sim1/3$ of the Earth's surface with a resolution of $\sim 10$ km at 500 frames per second in a narrow 1.1 nm pass band centered around the O I oxygen triplet at 777.4 nm \cite{goodman2013goes}. GLM is very efficient at detecting bright fireballs which usually saturate ground-based cameras \cite{jenniskens2018detection}. As camera saturation prevents an accurate estimate of meteoroid properties during atmospheric entry, GLM extends the usable measurement size range of bolides as compared to ground-based cameras. The larger ground-based camera coverage, which provides observations of fireball trajectories and orbits when fused with space-based light curves, records larger numbers of the decimeter-sized meteoroid population than previously possible and allows accurate estimates of their physical properties.

Observations from earlier fireball networks \cite{ceplecha1988earth, flynn2018physical} have established that decimeter-sized chondritic-like meteoroids which penetrate deeper into the atmosphere predominantly come from asteroidal low-inclination orbits. Similarly, most friable meteoroids which disrupt high in the atmosphere were measured to be on Jupiter-family comet (JFC), Halley-type comet (HTC), or long-period comet (LPC) orbits \cite{vojavcek2019properties}. Minor cross-contamination of material between asteroidal and JFC orbits is observed and can be explained by their dynamical evolution \cite{binzel2015near}, consistent with telescopic observations of comets and asteroids \cite{weissman2002evolution}. In-situ measurements have found rocky (refractory) materials in comets \cite{brownlee2012overview}, but these are small, microscopic chondrules and CAI fragments, presumed to be embedded during comet formation. However, the presence of macroscopic (dm-sized) rocky material originating from the Oort cloud (HTCs and LPCs) is much harder to explain. The abundance in the Oort cloud of larger, refractory material which likely formed in the inner Solar System would be a key diagnostic in distinguishing between dynamical models of early Solar System formation \cite{weissman1997origin, shannon2015eight, meech2016inner}.

Contemporary models which account for planetesimal collisions predict that a significant number of rocky objects can only be implanted in the Oort cloud during a Grand Tack dynamical instability episode caused by the radial migration of the giant planets early in Solar System history \cite{shannon2019oort}. The dynamical instability causes removal of $50-90\%$ of rocky material in what is now the asteroid belt in such a way as to reproduce the main-belt's observed orbital and compositional distribution \cite{walsh2011low, meech2020origin}. These migration models necessitate a ``massive'' proto-asteroid belt scenario and predict the ratio of icy to rocky planetesimals in the Oort cloud between 100:1 and 2000:1 \cite{weissman1997origin, izidoro2013compound, shannon2015eight, meech2016inner, raymond2018migration, shannon2019oort, zwart2021oort}. Following this early and fast migration, a slower dynamical diffusion process is postulated to further remove $\sim 70\%$ of the main-belt leaving the mass we see today \cite{bottke2005fossilized, minton2010dynamical}. Recent studies show that to match the dynamical and geochemical evidence, the instability occurs as early as $30-60$ Myr after the dissipation of gas in the protoplanetary disk \cite{clement2019excitation, de2020dynamical, marrocchiastrophysical}. 

% The key difference between various dynamical models of the formation of the Solar System is the initial amount of material in this proto-asteroid belt. 

The competing model of early Solar System formation, the pebble-accretion model \cite{bitsch2015growth}, eschews the migration scenario but allows for the rapid formation of the giant planets before the solar gas nebula dissipates \cite{morbidelli2016challenges}. In the pebble-accretion model, filaments of mm$-$ to cm-sized pebbles gravitationally collapse quickly to form planetesimals \cite{johansen2007rapid}. The planetesimals in the terrestrial region grow more efficiently than those beyond 1 au, so there is no requirement to scatter rocky material to explain the small masses of Mars and the asteroid belt \cite{levison2015growing}. As the initial mass of the proto-asteroid belt is assumed to be small, the pebble-accretion model predicts that virtually no scattered rocky objects are implanted into the Oort cloud. It predicts an icy/rocky ratio of at least 10,000:1 \cite{meech2016inner}.

The initial size of the proto-asteroid belt remains a contentious topic. Recent work \cite{raymond2017empty} suggests that the asteroid belt could have initially been empty and later populated separately by S- and C-type objects. In this model, S-type asteroids are implanted through simple gravitational diffusion as by-products of terrestrial planet formation with C-types implanted during the growth of the giant planets through aerodynamic drag destabilization \cite{raymond2017origin}. This scenario is also compatible with an early onset of dynamical instability and migration of the giant planets, which seems to be a necessary element in the reproduction of observational constraints \cite{clement2019early}.

Recently, it has been shown that the migration-induced dynamical instability is also compatible with an enhanced version of the pebble accretion model which uses realistic opacities \cite{broz2021early}. In this model, the terrestrial planets form fast in only $\sim10$ Myr, and then planet migration is invoked as one of the possibilities to explain the hafnium-tungsten anomaly in the Earth's mantle caused by the Theia impact \cite{desouza2021can}. Nevertheless, even though the 1.5 to 4 au region is assumed to be a divergence zone \cite{bitsch2014stellar} (in between the terrestrial and giant-planet convergence zones where growth occurs), planetesimals in the divergence zone only slowly dissipate into neighbouring regions and are not scattered into highly excited orbits.

% - can the scattered objects be expected to be small because they didn't get enough time to grow? perhaps our discovery can only constrain the sizes of smaller objects - the Batygin & Laughlin 2015 paper limits the size to 100 m but it seems to ignore many important things

The earliest evidence of macroscopic asteroidal material in the Oort cloud was the discovery of asteroid 1996 PW \cite{weissman1997origin}. Despite its highly eccentric orbit (eccentricity $e = 0.9907$, inclination $i = 30.09$, semi-major axis $a = 269.5~\mathrm{au}$, and period $p = 4424$ yr, solution date \href{https://ssd.jpl.nasa.gov/tools/sbdb_lookup.html\#/?sstr=1996\%20PW}{2021-Apr-14}), it showed no cometary-like activity. It had a D-type asteroid reflectance spectrum \cite{hicks2000physical, lamy2009colors, demeo2015compositional}, also similar to bare cometary nuclei observed at large solar distances; hence its origin as asteroid or an extinct cometary nucleus was uncertain.

More such tailless comet ``Manx'' objects \cite{meech2014c} have since been discovered, having a wide variety of surface properties, including S-type spectra consistent with anhydrous rocky material \cite{stephens2017chasing,  piro2021characterizing}. For example, comet C/2014 S3 (PANSTARRS) has an S-type reflectance spectrum, however it shows activity consistent with sublimation of water ice \cite{meech2016inner}. To explain the discovery of Oort cloud S-type objects in the context of the Solar System formation models, dynamical simulations of the evolution of the distribution of cometary and asteroidal objects were performed by ref. \cite{shannon2019oort}. Taking the collisional evolution of asteroids into account, they found that the Grand Tack model \cite{walsh2011low} is the only model which predicts a sufficient icy/rocky mass ratio of Oort cloud objects (on the order of 100:1) to explain the detection of the comet C/2014 S3.

Pan-STARRS1 observations of LPCs have found that there is a deficit  of objects with diameters $D \lesssim 1$ km \cite{boe2019orbit}, assuming the physical processes producing cometary activity are size-independent. These data show a significant change in the cumulative size-frequency distribution (SFD), where N$_{cum}$ $\propto$ D$^{-\alpha}$, for $D \sim 2.8$ km, with $\alpha = 3.6$ for larger, and $\alpha = 0.5$ for smaller objects, consistent with comet formation models \cite{davidsson2016primordial}. Pan-STARRS1 is able to detect LPC objects  down to a size of $D \sim 100$ m\cite{boe2019orbit}. This either means that the small objects are devoid of volatiles or they do not exist. 

% Further analysis \cite{boe2019distinguishing} has shown that $\sim 1/3$ of all objects on LPC orbits detected by Pan-STARRS1 do not show any activity (called tailless ``Manx'' objects), a surprising result considering the massive decrease in visibility of non-active bodies.

As telescopic measurements of an object in an LPC orbit may be compromised by space weathering \cite{binzel2004observed, kaluna2016space}, a more direct way to probe bulk physical properties of LPC material is desirable. One alternate method is to observe fireballs associated with an LPC-type meteoroid entering the atmosphere \cite{ceplecha1976fireball, borovivcka2020two}. Nevertheless, such observations suffer from small atmospheric collection areas, so detection of decimeter-sized objects on long-period comet orbits is rare \cite{brown2002flux}. 

By using the observed light curves and dynamics of mm$-$ and cm-sized cometary meteoroids, their ablation behaviour is well explained if they are modelled as a highly porous ($\sim90\%$) collection of $10-300~\mu$m sized silicate grains \cite{borovivcka2007atmospheric, hulfeld2021three}. The grain size distribution derived from observations of cometary meteoroids matches well to in-situ measurements of the JFC comet 67P/Churyumov–Gerasimenko  \cite{vojavcek2019properties, hornung2016first}. Similarly, observations of dm-sized meteorite-dropping fireballs can be well explained by modelling them as rocky objects which fragment deep in the atmosphere when aerodynamic loading exceeds their global mechanical strength \cite{borovivcka2013kovsice}. In most cases, global strengths of meteorite-producing fireballs are found to be much lower than the compressive strengths of their associated meteorites, a finding ascribed to internal cracks \cite{borovivcka2020two}.

Categorizing meteoroid strength relies on a relative comparison of atmospheric ablation behaviour. Generally speaking, slower, more massive, and stronger meteoroids penetrate deeper into the atmosphere. The PE criterion \cite{ceplecha1976fireball} removes the speed and mass bias so that the strength can be directly compared among observed meteoroids (see Methods). Meteoroids can be sorted into several groups based on material strength: Type I fireballs having PE $> -4.6$ are related to ordinary chondrites (strongest material), Type II with $-5.25 \leq $ PE $\leq -4.6$ are related to carbonaceous chondrites, while Type III with PE $< -5.25$ are cometary (weakest material).

The first observation of a multi-cm sized rocky meteoroid on an HTC orbit was recorded in 1997 over the Czech Republic, called the Karl\v{s}tejn fireball \cite{spurny1999en010697, spurny1999detection}. The $\sim 30$ gram object was on a retrograde orbit, $a = 3.5$ au, $i = 138^{\circ}$, $e = 0.7$, and $T_J = 0.62$ (where $T_J > 3$ are asteroidal orbits, $2 < T_J < 3$ are short period comet orbits, and $T_J < 2$ are LPC orbits). It entered the atmosphere at $65~\mathrm{km~s^{-1}}$ and penetrated down to an end height of 65~km, about 25~km deeper than cometary objects of similar speed and mass. It was classified as a Type I (rocky) fireball based on its PE value. Its spectrum was highly depleted in volatile elements, notably sodium, and distinct from cometary fireballs. It reached a maximum dynamic pressure ($P_{\mathrm{dyn}} = v^2 \rho_{\mathrm{air}}$) of 660 kPa before ablating away gradually, indicating that the true mechanical strength of the body was not reached as there was no evidence of catastrophic disruption. The dynamics of the body were consistent with a bulk density of 3700 $\mathrm{kg~m^{-3}}$. Nevertheless, the semi-major axis was smaller than most long-period comets and the dynamic pressures lower than what decimeter-sized chondritic meteorite-dropping fireballs survive, likely due to the small mass of the body. The authors theorized it was a cm-sized rocky component originally embedded in a comet, perhaps as part of an irradiated crust \cite{spurny1999en010697}.

Records from decades of meteor shower observations have not revealed any macroscopic ($>$ cm-sized) lithic material mixed in with fragile HTC or LPC meteoroids. Smaller inclusions have been documented, notably several mm$-$sized Type I fragments of Leonid fireballs were observed during the 1998 Leonid fireball storm \cite{borovivcka2000time, kokhirova2011observations, borovicka2019physical} as have some gram-sized Type I Taurids \cite{spurny2017discovery}. However, the Taurids are an unusual stream; they are on the dynamical boundary between JFC and asteroidal orbits, are generally classified as Type II material, can be difficult to separate from the sporadic background, and have an origin likely related to fragmentation rather than gas drag sublimation \cite{borovivcka2020physical}.

\newpage
\section*{Results}
% - concise, focused account of the findings (headed 'Results')
% - introduce results + modelling results
% - 5x paragraphs = interpretation + more results (e.g. flux)

\begin{figure}[!ht]
	\begin{center}
 		\includegraphics[width=\textwidth]{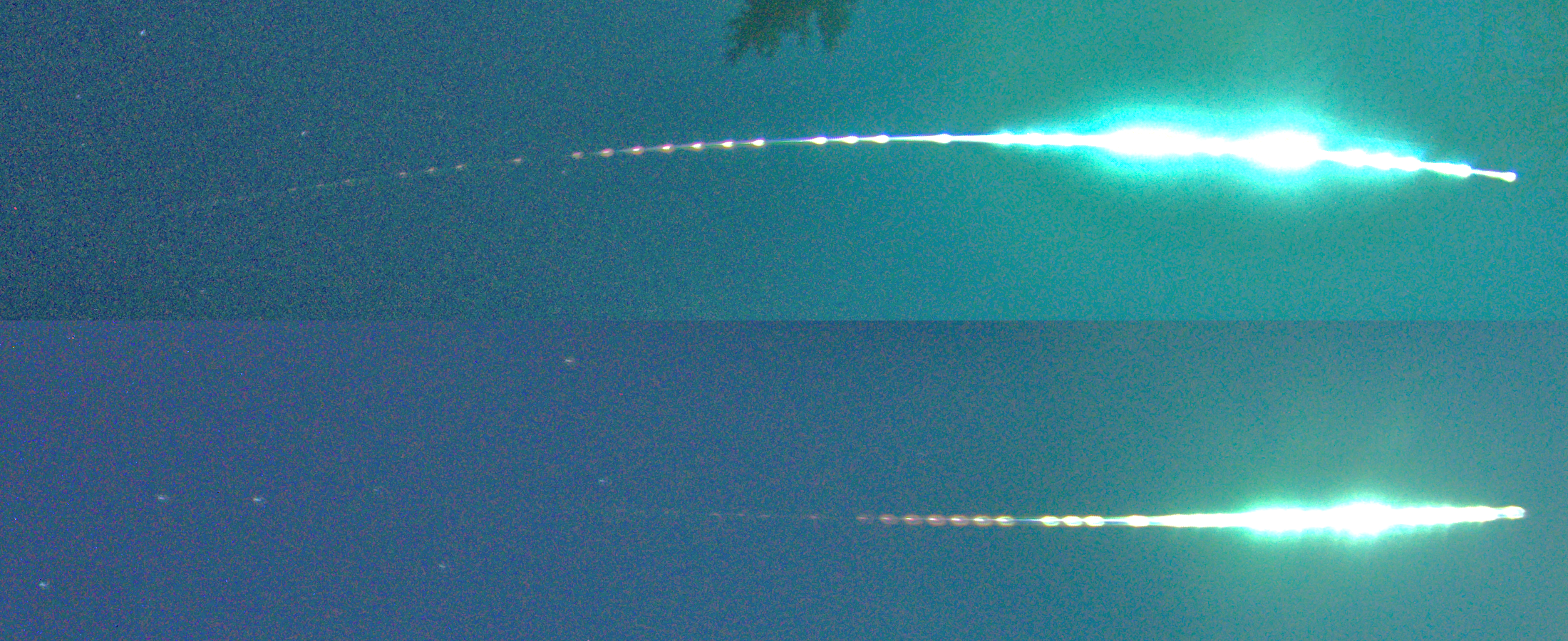}
	\end{center}
	\caption{The fireball as seen from the two GFO stations. It was observed for a total of 2.4~s with a path length of 148.5~km. Top: Miquelon Lake. Bottom: Vermilion (the Big Dipper can be seen at the left side of the inset). The fireball is moving left to right, and the periodic breaks in the fireball are used to encode the absolute time to an accuracy of 1~ms.}
	\label{fig:gfo_images}
\end{figure}

Here we report the direct observation of a decimeter-sized rocky meteoroid (PE = -4.49, Type I) on a long-period comet orbit (i = $121^{\circ}$, $e \approx 1.0$, $T_J = -0.46$, see Table \ref{tab:orbit}). This meteoroid reached dynamic pressures similar to those of ordinary chondrites. The $\sim 2$~kg body entered the atmosphere $\sim 100$~km north of Edmonton, Alberta, Canada on February 22, 2021 at 13:23:17 UTC. Its full atmospheric luminous path was recorded by two Global Fireball Observatory (GFO) all-sky cameras \cite{devillepoix2020global} (see Figure \ref{fig:gfo_images}) and over 200 security and dash cameras. In addition, it was detected by the Geostationary Lightning Mapper (GLM) instruments on board the Geostationary Operational Environment Satellites (GOES) 16 and 17, permitting measurement of its unsaturated light curve (see the Photometric Calibration section in Supplementary Information). We used the most recent astrometric calibration methods \cite{vida2021global} and computed the atmospheric trajectory (internal accuracy of 30~m) using the GFO data and using one additional security camera (accuracy 70~m) (see the Astrometric Calibration section in Supplementary Information). The fireball entered the atmosphere with a velocity of $62.1~\mathrm{km~s^{-1}}$ and penetrated down to a height of 46.5~km, about 20~km deeper than the Karl\v{s}tejn event that had a similar velocity but a $70\times$ smaller mass. The parent body search did not return any matches, an expected result given the large orbital period.

\begin{table}[!ht]
\caption{Geocentric radiant and heliocentric orbit (J2000.0).}
\label{tab:orbit}
\begin{tabular}{llrrr}
Description & & Nominal value & \multicolumn{2}{c}{95\% Confidence Interval} \\
            & &               & Lower & Upper \\
\hline\hline % inserts double horizontal lines
Geocentric right ascension of radiant         & $\alpha_g$ & $ 271.922^{\circ}$    & $271.856^{\circ}$    & $271.990^{\circ}$ \\
Geocentric declination of radiant             & $\delta_g$ & $  +4.40^{\circ}$     & $+4.19^{\circ}$      &  $+4.64^{\circ}$ \\
Geocentric velocity ($\mathrm{km~s^{-1}}$) & $v_{g}$     & $  60.97       $     & 60.89                &  61.00 \\
Semimajor axis (au)                           & $a$        & $ 104           $     & 50                   & 230 \\
Eccentricity                                  & $e$        & $   0.9941      $     & 0.9878               &   0.9973 \\
Perihelion distance (au)                      & $q$        & $   0.6150      $     & 0.6126               &   0.6170 \\
Argument of perihelion                        & $\omega$   & $ 103.95^{\circ}$     & $103.51^{\circ}$     & $104.25^{\circ}$ \\
Longitude of ascending node                   & $\Omega$   & $ 333.857472^{\circ}$ & $333.857460^{\circ}$ & $333.857485^{\circ}$ \\
Inclination                                   & $i$        & $ 121.40^{\circ}$     & $120.98^{\circ}$     & $121.80^{\circ}$ \\
Aphelion distance (au)                        & $Q$        & $ 207           $     & 100                  & 459 \\
Period (yr)                                   & $T$        & $ 1059          $     & 357                  & 3484 \\
Last perihelion date                          &            & 2021-01-16.23         & 2021-01-16.06        & 2021-01-16.33 \\
Tisserand parameter w.r.t. Jupiter            & $T_J$      & $  -0.46        $     & -0.48                &  -0.40 \\
\end{tabular}
\end{table}

\subsection*{Ablation modelling}

The dynamics, light curve, and the fragmentation behaviour were modelled using a semi-empirical meteoroid ablation model \cite{borovivcka2013kovsice} (see Methods) which has been successfully applied to multiple meteorite-producing fireballs, as well as cometary meteoroids. The comparison between observations and the model fit is shown in Figure \ref{fig:model_fit}; the modelling details are given in the Modelling Results section in Supplementary Information. A bulk density of $\rho_{\mathrm{m}} = 3300~\mathrm{kg~m^{-3}}$, as appropriate for chondritic meteorite-dropping fireballs \cite{flynn2018physical}, fits the observed dynamics well.

\begin{figure}[!ht]
	\begin{center}
		\includegraphics[width=\textwidth]{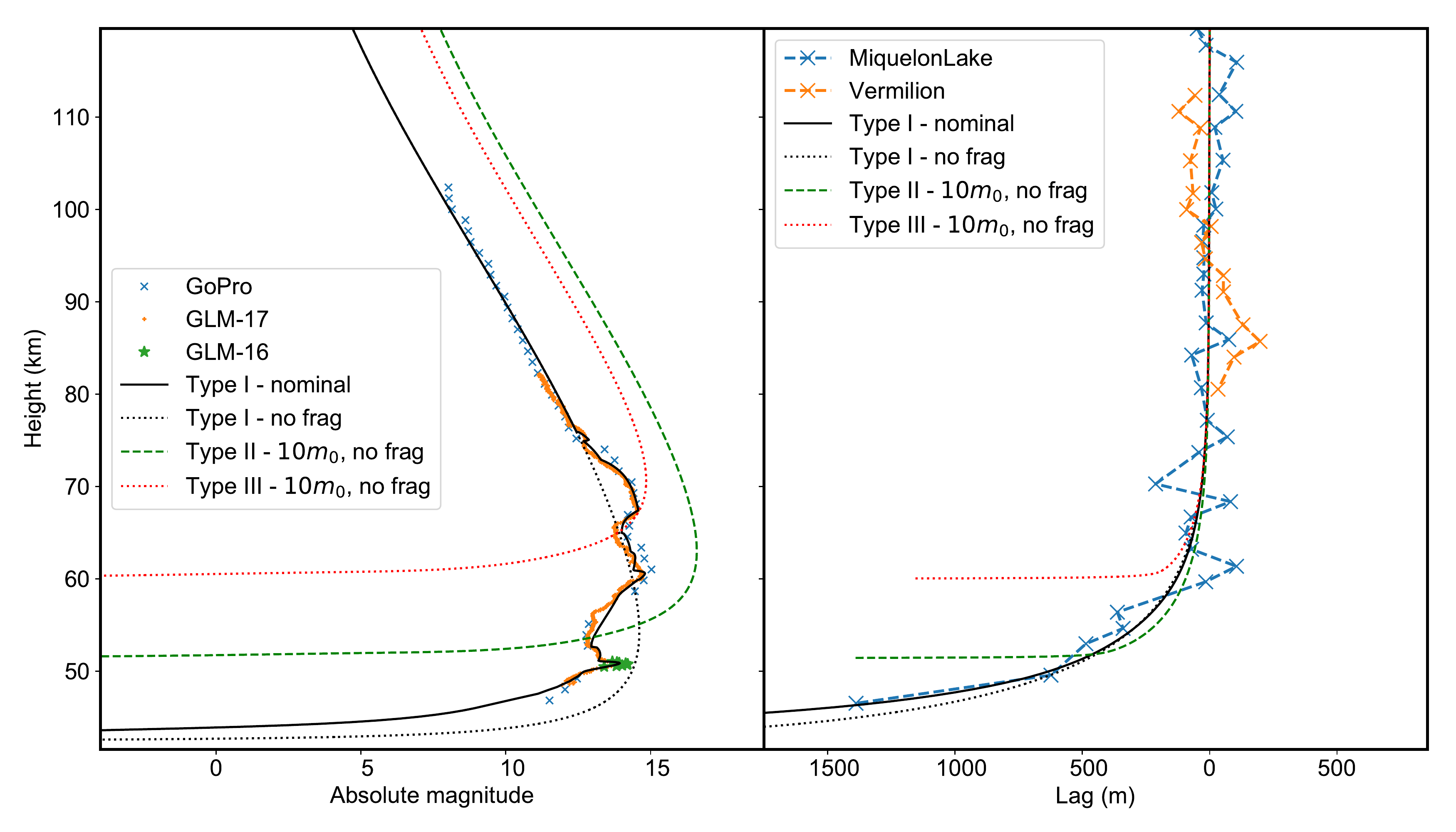}
	\end{center}
	\caption{Left: Observed and simulated light curve. The solid black line is the nominal fit with the parameters and fragmentation behaviour given in the Supplementary Materials (Supplementary Tables \ref{tab:model_phys} and \ref{tab:fragmentation}). A no-fragmentation solution for Type I (black), II (green), and III (red) objects is also shown. The GLM light curve was calibrated to an accuracy of $\pm 0.1$~mag using three independent high speed fireballs where GLM light curves and ground-based records were available. The GoPro light curve was calibrated to an accuracy of $\pm 0.3$~mag. Right: Observed and simulated deceleration profile (lag) for various simulation scenarios. The lag is the distance a decelerating meteoroid falls behind a hypothetical non-decelerating meteoroid moving at the initially observed velocity. As the fireball showed a wake, it was not possible to determine the along the trajectory positions of the leading fragment to the same  accuracy as the transverse positions. The lag measurement errors are reflected in the scatter around the zero lag axis which is on the order of $\pm 200$~m. The Cochrane security camera was not used for velocity measurement due to lower accuracy.}
	\label{fig:model_fit}
\end{figure}

As a demonstration of how improbable that this was a weak, cometary-like body, we modelled a non-fragmenting fireball using physical properties appropriate for cometary meteoroids (ablation coefficient $\sigma = 0.08~\mathrm{kg~MJ^{-1}}$, $\rho_{\mathrm{m}} = 1000~\mathrm{kg~m^{-3}}$, and luminous efficiency $\tau$ for Type III bodies \cite{ceplecha1998meteor, revelle2001bolide}), with a $10\times$ larger pre-atmospheric mass (20~kg) than the Alberta fireball keeping the same trajectory parameters. The hypothetical cometary meteoroid only penetrated down to a height of $\sim 60$~km. 

The simulation was repeated for a hypothetical carbonaceous chondrite meteoroid ($\sigma = 0.042~\mathrm{kg~MJ^{-1}}$, $\rho_{\mathrm{m}} = 2000~\mathrm{kg~m^{-3}}$, $\tau$ for Type II bodies \cite{ceplecha1998meteor, revelle2001bolide}) with the same 20~kg mass. In this case, the meteoroid only penetrated to a height of 53~km while the simulated light curve was $\sim 2.5$ magnitudes brighter than observed. 

We found no combination of model parameters for carbonaceous-chondrite or cometary-like material which could fit the observations. If these hypothetical Type II and III objects had the same mass as our Type I rocky object, their end heights would have been a further 8~km higher than these values. From other fireball measurements, a cometary body at these speeds and masses is not expected to withstand the aerodynamic loading below $\sim 80$~km without fragmenting \cite{borovivcka2020physical}.

The fireball fragmented under dynamic pressures similar to those observed for rocky meteoroids. Ref. \cite{borovivcka2020two} analyzed the fragmentation behaviour of several instrumentally observed ordinary chondrite (OC) meteorite dropping fireballs. They found that OC meteoroids do not fragment randomly but follow a specific pattern. The first phase of fragmentations occurs at dynamic pressures between 0.04–0.12~MPa, interpreted as being caused by detaching of weakly cemented fragments from the surface. The second phase occurs between 0.5-5~MPa, presumably due to weaknesses associated with internal cracks.

\begin{figure}[!ht]
	\begin{center}
		\includegraphics[width=0.75\textwidth]{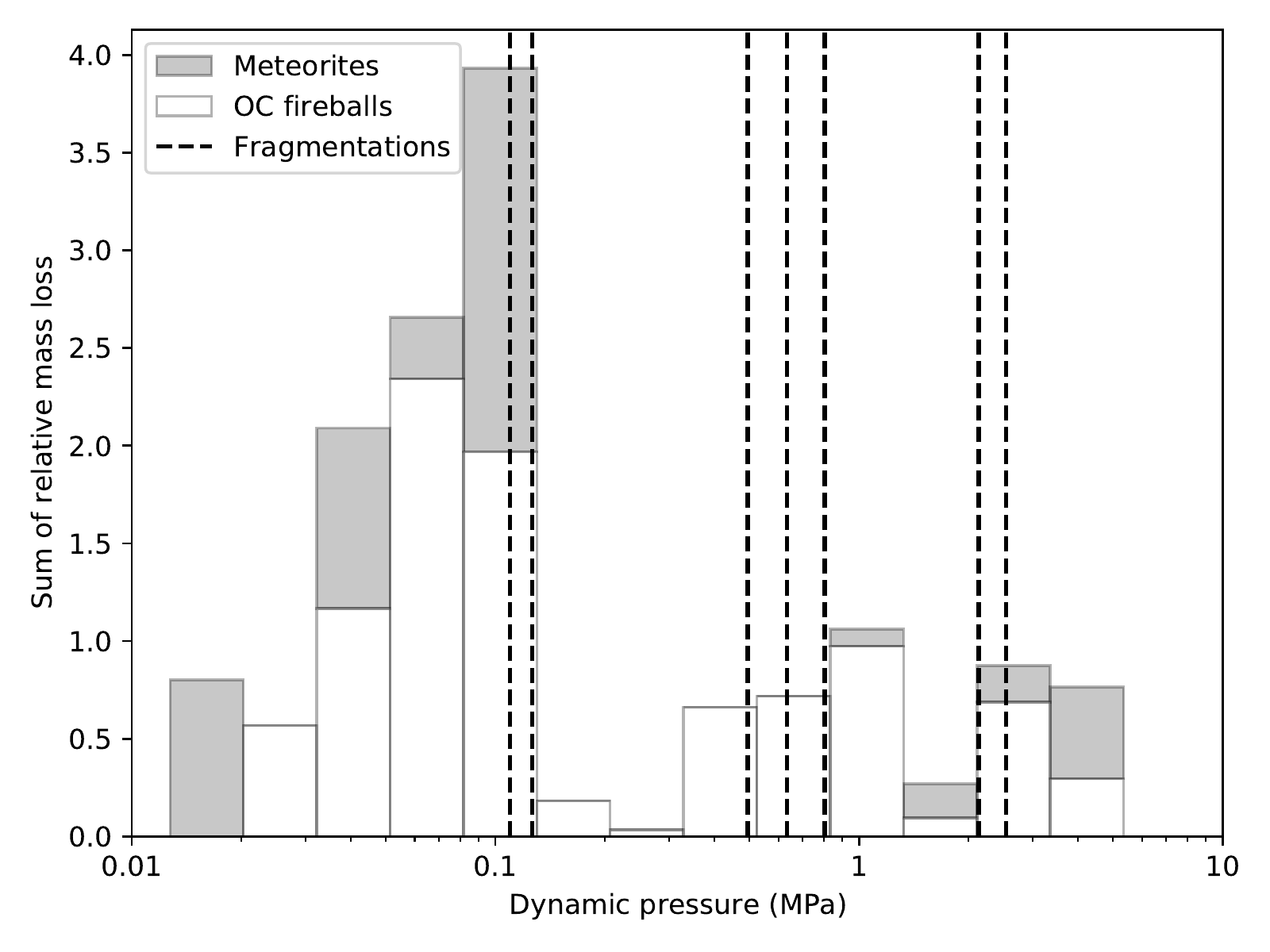}
	\end{center}
	\caption{Observed fragmentations of the Alberta meteoroid (vertical dashed lines) compared to previously observed fragmentation behaviour of ordinary chondrite (OC) fireballs. The dynamic pressure ($\Gamma \rho_{\mathrm{air}} v^2$) is computed using the drag coefficient of $\Gamma = 1.0$ to be consistent with values from ref. \cite{borovivcka2020two}. The uncertainty in the dynamic pressure due to short-term variations in the air mass density which are not captured by the NRLMSISE-00 model \cite{vida2021high} is about 25\%. However, uncertainties due to the unknown values of $\Gamma$ in the flow regime of this meteoroid are not quantified. The gray bars show the sum of relative mass loss at the given dynamic pressure for fireballs with recovered meteorites, and white bars are for (OC) fireballs with currently unrecovered meteorites, or which were too small to produce meteorites}.
	\label{fig:strength_comparison}
\end{figure}

Figure \ref{fig:strength_comparison} shows their measurements of relative fragmentation mass loss versus the dynamic pressure for a collection of fireballs with recovered OC meteorites and fireballs of OC physical properties that were too small to produce meteorites. The dynamic pressure at fragmentation measured for the Alberta fireball matches well to the chondritic fragmentation profile.

% However, the body never catastrophically disrupted into constituent grains, even after reaching a maximum dynamic pressure of 2~MPa, a further indication that it was very strong.

\subsection*{Flux of rocky meteoroids from the Oort cloud}

It is possible to estimate the flux and the ratio of Type I (rocky) objects in comparison to weaker Type II/III from the Oort cloud at the limiting mass appropriate for fast fireballs. The only published source for which there are reliable mass and time-area product estimates is the fireball data set of the Meteorite Observation and Recovery Project (MORP)\cite{halliday1996detailed}. MORP is the the only unbiased ``clear sky'' survey of fireballs. There are a total of 30 fireballs with $T_J < 2$, speed $> 50~\mathrm{km~s^{-1}}$, and a mass larger than 10~g in the MORP data set. The speed limit removes the speed-dependent mass sensitivity \cite{ceplecha1998meteor}, allowing us to set consistent limiting mass for the flux. We set the mass limit to 10~g (in the contemporary mass scale following ref. \cite{borovivcka2020two}), for which the data set is complete. This mass is also significantly larger than any previously observed refractory inclusions in cometary material; the largest Type I Taurids are an order of magnitude smaller \cite{spurny2017discovery}.

\begin{figure}[!ht]
	\begin{center}
		\includegraphics[width=0.75\textwidth]{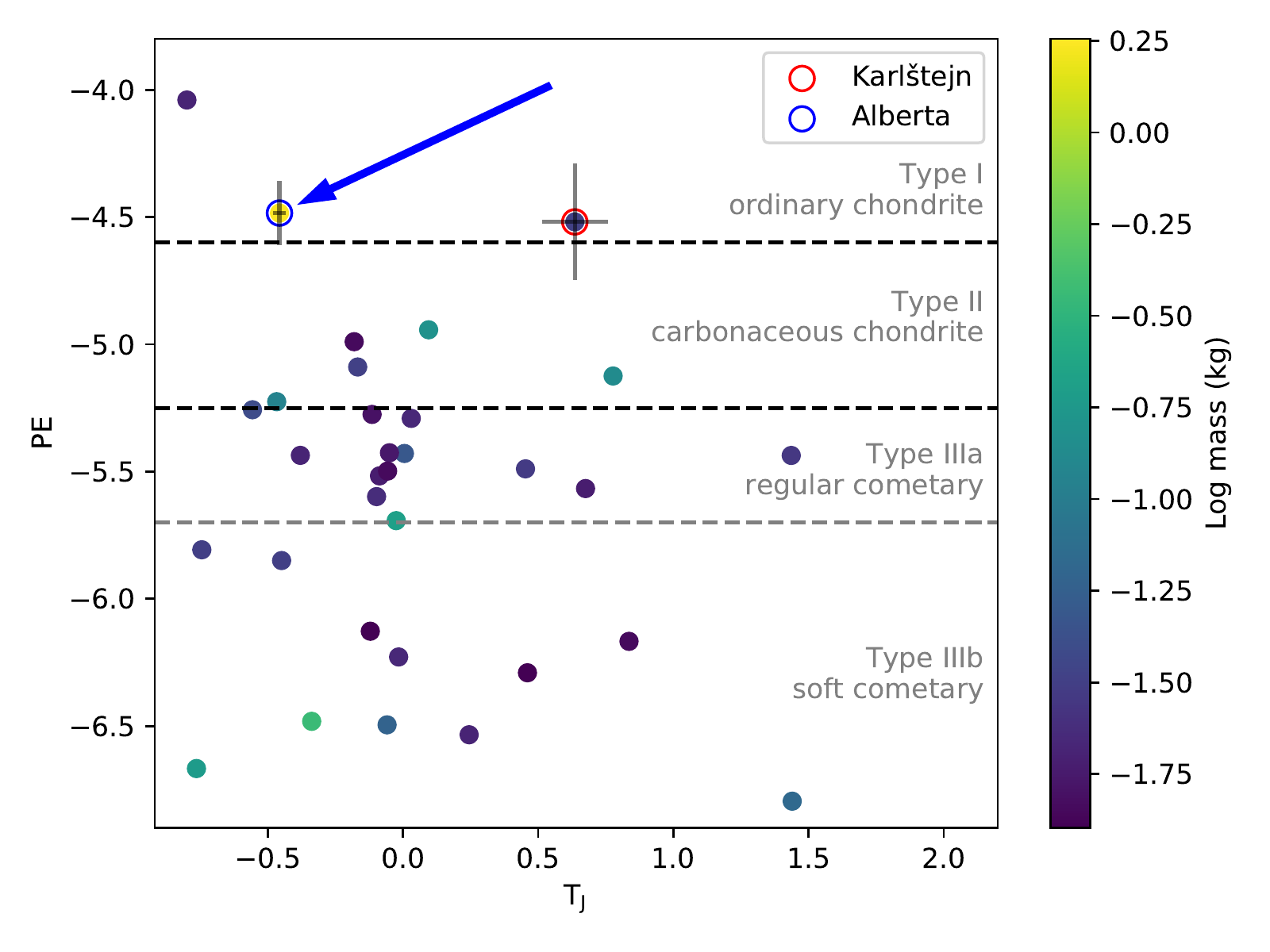}
	\end{center}
	\caption{All published fireball data showing PE as a function of Tisserand's parameter ($T_J$) with respect to Jupiter for $T_J < 2$ fireballs in the MORP data set \cite{halliday1996detailed}. PE is a meteoroid strength factor, see the Methods section for more details. The Type I (rocky) objects of interest above the PE = -4.6 dividing line between Type I/II fireballs form a clearly separate grouping. The blue arrow points to the Alberta event. The MORP data set does not contain formal uncertainties, but they are of the same order as for the Alberta event.  % Note that the Alberta fireball is the largest mass fireball of any type recorded on an LPC orbit to date.
	}
	\label{fig:tj_vs_pe}
\end{figure}

Among the population of 30 fireballs with $T_J < 2$, only one is a Type I object (MORP catalog \#441 with $m = 20$~g), which was also on a retrograde orbit ($e = 0.969$, $i = 159.7^{\circ}$). This additional fireball, together with Karl\v{s}tejn and the Alberta event, forms an isolated group among all published fireballs: they have $T_J < 1$, retrograde orbits, and are quite far from the PE dividing line for Type II objects (see Figure \ref{fig:tj_vs_pe} and Table \ref{tab:typeI_objects}).

\begin{landscape}
\begin{table*}
\caption{Table of all $T_J < 2$ Type I objects in the literature satisfying our criteria. The errors for Karl\v{s}tejn and the Alberta event are the 95\% confidence intervals. We assumed a conservative uncertainty in the mass of the Alberta event of a factor of 2, due to possible variations in the luminous efficiency \cite{subasinghe2017luminous}. Karl\v{s}tejn has larger uncertainties as the mass was only derived dynamically (0.003 to 0.1~kg), which we converted to the contemporary mass scale (from ref. \cite{ceplecha1976fireball} to ref. \cite{borovivcka2020two}). The orbital uncertainties for the Karl\v{s}tejn fireball are symmetric and taken from the original publication\cite{spurny1999en010697}. The MORP data set does not contain any uncertainties. The uncertainty of the ascending node is near zero as it reflects the time of the meteoroid intersecting Earth's orbit (i.e. entering the atmosphere) which is accurately known. However for MORP, the uncertainty in the timing for some fireballs is on the order of tens of minutes ($\pm 10^{-5\circ}$ in the ascending node).
}
\label{tab:typeI_objects}
\begin{tabular}{lrrlllllrr}
Name                                   & PE    & $T_J$ & Mass   & a     & q     & e     & i      & $\omega$ & $\Omega$ \\
                                       &       &       & (kg)   & (au)  & (au)  &       & (deg)  & (deg)    & (deg) \\
\hline\hline % inserts double horizontal lines
Alberta (this work)                    & $-4.49^{+0.12}_{-0.13}$ & $-0.46^{+0.06}_{-0.02}$ & $1.8^{+1.8}_{-0.9}$    & $104^{+126}_{-54}$   & $0.615^{+0.002}_{-0.002}$ & $0.994^{+0.003}_{-0.006}$ & $121.40^{+0.4}_{-0.4}$ & $104.0^{+0.3}_{-0.4}$    & 333.86 \\
Karl\v{s}tejn\cite{spurny1999en010697} & $-4.53^{+0.42}_{-0.23}$ &  $0.64^{+0.12}_{-0.18}$ & $0.033^{+0.082}_{-0.030}$ & $3.5^{+0.18}_{-0.18}$ & $1.012^{+0.0002}_{-0.0002}$ & $0.710^{+0.016}_{-0.016}$ & $137.90^{+0.10}_{-0.10}$ & $174.60^{+0.14}_{-0.14}$    &  71.55 \\
%MORP \#434\cite{halliday1996detailed}  & -4.54 &  0.77 &  0.005 &   5.5 & 0.236 & 0.957 & 107.6  & 303.7    & 256.4  \\
MORP \#441\cite{halliday1996detailed}  & -4.04 & -0.78 &  0.020 &  24.7 & 0.765 & 0.969 & 159.7  & 56.7     & 103.8
\end{tabular}
\end{table*}
\end{landscape}

The total time-area product for the MORP clear-sky survey was $1.51 \times 10^{10}$~$\mathrm{km^2~h}$ \cite{halliday1996detailed}. Given the 29 MORP fireballs satisfying our orbital criteria, the total flux of carbonaceous and cometary (Type II and III) objects can be computed. Assuming that times of meteor events are distributed according to the Poisson distribution \cite{vida2020new}, the 95\% confidence interval is computed using the exact method \cite{ulm1990simple} as $[\mathrm{Pr}(\alpha/2, 2x), \mathrm{Pr}(1-\alpha/2, 2(x + 1))]$, where $\mathrm{Pr}$ is the percent point function of the $\chi^2$ distribution, $x = 29$ is the number of observed events, and $\alpha=0.05$ (i.e. the 95\% confidence interval). The corresponding range of observed events is $[19.4, 41.7]$, producing a flux of $16.8^{+7.3}_{-5.6}~\mathrm{meteoroids/10^6 km^2/yr}$ to a mass limit of 10~g.

% Source of the Poission method: http://onbiostatistics.blogspot.com/2014/03/computing-confidence-interval-for.html

As a check on the flux values and ranges we derive, we use data from the Global Fireball Observatory cameras in Alberta. These were gradually deployed starting with three cameras in July 2018, completing deployment with five cameras in November 2018. We used known deployment times and atmospheric coverage to compute a time-series of common collecting area for the network; $0.24 \times 10^{6}$~$\mathrm{km^2}$ for the three and $0.45 \times 10^{6}$~$\mathrm{km^2}$ for the five camera configuration. The common area was computed as an intersection of at least two camera fields of view at the height of 70~km and within a radius of 300~km from an individual camera \cite{devillepoix2020global}. Given that the average daily effective collection time is $3~\mathrm{h/day}$ in Alberta \cite{oberst1998european}, we estimate that the time-area product for the network up until February 2021 is $1.2 \times 10^{9}$~$\mathrm{km^2~h}$. This network has yet to record a single Type II or III object on an LPC orbit in that time period, although it should have observed $\sim2$ given the MORP flux. Using Poisson statistics, the probability for this non-observation is 10\%, which is within statistical significance.

The total time-area product for $T_J < 2$ rocky Type I events is $1.63 \times 10^{10}$~$\mathrm{km^2~h}$, derived by combining the values from the two networks. Note that the product is entirely dominated by the MORP survey, making up 90\% of the total. We only include the Alberta event and the MORP \#441 fireball in the flux estimate. To include the Karl\v{s}tejn fireball we would need an accurate estimate of the completeness and the time-area product for the contemporary European Network \cite{oberst1998european}, which is not available.

Given that only two events were observed, the 95\% Poisson confidence interval is $[0.24, 7.22]$ for the number of events. The total flux of rocky Type I objects on LPC orbits is thus $1.08^{+2.81}_{-0.95}~\mathrm{meteoroids/10^6~km^2/yr}$ for a mass limit of 10~g, or $6.0^{+12.8}_{-5.2}$\% of the total flux of all meteoroids impacting Earth on LPC orbits. This suggests of order 1--20\% (about one in five to about one in a hundred) of LPC meteoroids at tens of gram sizes or larger are rocky.

\newpage
\section*{Discussion}
% - 1 or 2 short paragraphs
% - position results within the big picture
% - last paragraph: most impactful conclusion and what difference it makes for the general understanding

The confirmation of the existence and a comparatively high abundance of macroscopic lithic objects in the Oort cloud, constraining the ratio of icy/rocky objects to between 130:1 and 5:1 for masses $>10$~g (95\% CI), supports the need for a mechanism of ejection of inner Solar System material into nearly hyperbolic orbits. Even in a scenario where most of the Oort cloud objects are captured from other star systems \cite{levison2010capture}, an ejection mechanism still needs to be present to explain the radial mixing of material. 

We interpret the icy/rocky ratio as an intrinsic parameter of the population, assuming that it has remained unchanged in the Oort cloud since its implantation during the formation of the Solar System, as it is consistent with reflectance spectra surveys of large Oort cloud objects \cite{shannon2019oort} which predict an icy/rocky ratio on the order of 100:1.

Was the Alberta fireball itself a primordial object? Collisions between similarly-sized Oort cloud objects are very rare across the range of sizes \cite{stern1988collisions, stern2003evolution}, however $>100$~m bodies are known to experience surface processing due to impacts of m-sized and smaller objects. Surface gardening and erosion due to $\mathrm{\mu}$m-sized dust impacts from the interstellar medium (ISM) \cite{weissman1996oort} is generally experienced by objects of all sizes. The ISM erosion model predicts that all primordial objects smaller than a few meters should have been eroded away \cite{stern1990ism}, indicating that the Alberta fireball possibly originated from a larger parent asteroid.

% NO LONGER APPLIES AFTER THE MASS OF KARLSTEJN WAS CONVERTED TO THE MODERN SCALE:
% We note that the size distribution of the observed cometary and rocky fireballs on LPC orbits differs significantly. Cometary meteoroids are mostly observed at small sizes ($<50$~g), while the only observed rocky meteoroids have masses of 20, 30, and $\sim2000$~g. This is consistent with the usual interpretation that cometary material is fragile and falls apart over time \cite{jewitt2004cradle, jenniskens2017meteor}. However, assuming a classical power-law mass distribution, it is not clear why smaller rocky meteoroids are infrequently observed.

Our findings support a massive proto-asteroid belt scenario as the source of rocky objects. Recent work \cite{shannon2019oort} shows that the Grand Tack dynamical instability model is the only one able the reproduce the observed abundance of rocky material in the Oort cloud predicting that the total icy/rocky ratio depends on the duration of the instability. As the model predicts more rocky material is implanted if the instability occurs faster, direct measurement of the icy/rocky fraction can be used to constrain the duration of the early instability. These findings challenge Solar System formation models based on pebble accretion alone, which currently cannot explain the high observed abundance of rocky material in the Oort cloud as derived from fireball measurements and telescopic reflectance spectra data.
% the presence of any significant icy/rocky ratio. The most recent iteration of the model \cite{izidoro2021planetesimal} successfully reproduces the orbital and compositional structure of the inner and outer Solar System from a set of dust rings. However all pebble accretion models so far have not explained

% $>100$ kPa …  and the main breakup occurred under loads exceeding 1 MPa

% Unlike telescopic observations which rely on reflectance spectra (and so are compromised by space weathering), our inference is more direct as we are actually probing the true physical strength of the object. 

% *** METHODS ***
\newpage
\section*{Methods}

\subsection*{Ablation and fragmentation model} \label{subsec:ablation_modelling}
The dynamics and light curve of the fireball were simulated using the established semi-empirical model \cite{borovivcka2013kovsice} which was successfully applied to reconstruct the fragmentation behaviour and physical properties of many meteorite dropping fireballs \cite{borovivcka2015instrumentally, borovivcka2017january, borovivcka2019maribo, borovivcka2020two} and fainter meteors \cite{borovivcka2007atmospheric, vojavcek2019properties}. In this model, the meteoroid is initially treated as a single body, but increases in brightness and sudden deceleration are explained by fragmentation. Previous works established several main modes of fragmentation: splitting into several single-body fragments, steady erosion of 10~$\mu$m~$-$~1~mm sized refractory constituent grains from the meteoroid's surface, ejection of an eroding fragment, and a sudden release of dust (i.e. a large number of constituent grains)\cite{borovivcka2013kovsice}.

All fragments and grains are modelled using the classical equations of single-body ablation \cite{ceplecha1998meteor}:

\begin{equation} \label{eq:deceleration}
    \frac{dv}{dt} = -K m^{-1/3} \rho_{\mathrm{air}} v^2\,,
\end{equation}

\begin{equation} \label{eq:mass_loss}
    \frac{dm_a}{dt} = -K \sigma m^{2/3} \rho_{\mathrm{air}} v^3\,,
\end{equation}

\noindent where $K$ is the shape density coefficient, $m$ the meteoroid mass ($m_a$ is the ablated mass), $v$ the velocity, $\rho_{\mathrm{air}}$ the atmosphere bulk density (NRLMSISE-00 model \cite{picone2002nrlmsise}), and $\sigma$ is the ablation coefficient. The ablation coefficient regulates how much mass is removed from the meteoroid per unit energy, and is usually expressed in $\mathrm{kg~MJ^{-1}}$ ($\mathrm{s^2~km^{-2}}$ is also often found in the literature). The parameter $K$ is used because the meteoroid density and shape cannot be measured separately:

\begin{equation} \label{eq:shape_density_coeff}
K = \Gamma A \rho_{\mathrm{m}}^{-2/3} \,,
\end{equation}

\noindent where $\Gamma$ is the drag coefficient, $A$ is the shape coefficient (1.21 for spheres which we adopt), and $\rho_{\mathrm{m}}$ is the meteoroid (or grain) bulk density. The equations were numerically integrated using a $\mathrm{4^{th}}$ order Runge-Kutta method and a time step of 2~ms. The integration of individual fragments is stopped if their mass falls below $10^{-14}$~kg or the speed below $3~\mathrm{km~s^{-1}}$, which is the ablation limit \cite{ceplecha1998meteor}.

The luminosity produced by ablation is computed as:

\begin{equation}
    I = -\tau \frac{v^2}{2} \frac{dm_a}{dt}  + m v \frac{dv}{dt}\,,
\end{equation}

\noindent where $\tau$ is the luminous efficiency. In this work we use the modern luminous efficiency function of ref. \cite{borovivcka2020two} to model the observed event. 

If the meteoroid or an ejected fragment is set to erode at a given time in the model, the total mass lost in erosion is regulated by the erosion coefficient $\eta$ which is applied in the same manner as the ablation coefficient in equation \ref{eq:mass_loss}:

\begin{equation} \label{eq:mass_erosion}
    \frac{dm_e}{dt} = -K \eta m^{2/3} \rho_{\mathrm{air}} v^3\,.
\end{equation}

\noindent For grain bulk density, we use $3500~\mathrm{kg~m^{-3}}$, appropriate for refractory silicate grains. The total mass loss at a given time is then the sum of the ablation and erosion mass loss:

\begin{equation}
    \frac{dm}{dt} = \frac{dm_a}{dt} + \frac{dm_e}{dt} \,.
\end{equation}

Inspired by in-situ observations of the mass distribution of cometary dust \cite{hornung2016first}, the masses of eroded grains are distributed according to a power-law $n(m) \sim m^{-s}$ \cite{popova2019modelling} where $n(m)$ is the number of grains of a given mass $m$, and $s$ is the differential mass distribution index. An upper ($m_u$) and a lower ($m_l$) grain mass limit is set during the modelling. To speed up computation, this mass range is binned into $z$ bins per order of magnitude, thus the integration of ablation equations is done only once for every mass bin instead for every fragment. A mass sorting parameter can be defined as $p = 10^{1/z}$ and it follows that the total number of grain mass bins within a range of masses is $k = \lceil \log(m_l/m_u)/\log p \rceil$. The total number of grains having a mass equal to the upper mass limit is then:

\begin{equation}
    n_u = \left\{\begin{array}{ll}
        \frac{m_e}{k m_u}, & \text{for } s = 2 \, \\
        \frac{m_e}{m_u} \frac{1 - p^{(2 - s)}}{1 - p^{k(2-s)}}, & \text{for } s \neq 2 \,,
        \end{array} \right.
\end{equation}

\noindent where $m_e$ is the total eroded mass at the given time step. The mass of every bin $i$ is then $m_i = m_u p^i$ for $i = 0, 1, ..., k - 1$. The number of discrete grains $N_i$ in every bin can be computed as:

\begin{equation}
    n_i    = n_u (m_u/m_i)^{(s-1)} + \frac{\Delta m_{i - 1}}{m_i} \,,
\end{equation}

\begin{equation}
    N_i = \lfloor n_i \rfloor \,,
\end{equation}

\noindent where $\Delta m_i = m_i(n_i - N_i)$ is the leftover mass in the mass bin after making the number of grains discrete ($\Delta m_0 = 0$). The leftover mass from the larger mass bins is distributed into smaller mass bins to ensure that there is no "virtual" mass loss due to numerical rounding. The grains are then ablated as single bodies until exhaustion following equations \ref{eq:deceleration} and \ref{eq:mass_loss}. A separate luminosity $I_i$ is computed for every mass bin, and the total luminosity produced by all grains at a given time is simply $\sum_{i = 0}^{k - 1} N_i I_i$.

Finally, after all fragments and grains have been fully integrated and their masses depleted, the magnitude $M$ of the fireball as it would have be seen at a distance of 100~km is computed as:

\begin{equation}
    M = -2.5 \log \frac{I}{P_{\mathrm{0m}}} \,,
\end{equation}

\noindent where $P_{\mathrm{0m}}$ is the power that a meteor needs to radiate in the camera's spectral band-pass so that it has an apparent magnitude of $0$~mag at a range of 100~km. In this work, we use a value of 1300~W, as appropriate for a high-speed meteor in the spectral band-pass of CMOS/CCD sensors \cite{weryk2013simultaneous}.

\subsection*{Meteoroid strength and dynamic pressure}

From decades of observations, it is now well established that there is a strong correlation between meteoroid material type and bulk strength \cite{borovivcka2005physical}. Cometary meteoroids are weak and disrupt under dynamic pressures of $\sim$ 1~kPa \cite{ceplecha1998meteor} upon entering the atmosphere, while asteroidal meteoroids can withstand pressures of 100~kPa prior to any fragmentation \cite{borovivcka2020two}, with their strongest components withstanding pressures of up to $1-10$~MPa without catastrophic disruption \cite{borovicka1998bolides}. These differing strengths explain why cometary fireballs break up at heights above 70~km \cite{spurny2017discovery}, while meteorite dropping fireballs break up typically below 40~km \cite{borovivcka2020two}. For cometary meteoroids, the strength of constituent $10-300~\mu$m silicate grains is on the order of 10s of MPa \cite{brownlee2006comet}, but high porosity significantly reduces the strength of larger grain aggregates \cite{kimura2020tensile}. 

Meteoroid fragmentation is commonly assumed to occur when the dynamic pressure $P_{\mathrm{dyn}} = \Gamma \rho_{\mathrm{air}} v^2$ exceeds the mechanical strength of the body \cite{popova2019modelling}. In most cases, meteorite-dropping fireballs of asteroidal origin fragment in two phases: the first from 0.04–0.12 MPa and the second from 0.5–5 MPa \cite{borovivcka2020two}. These are significantly lower than the tensile strengths of ordinary chondrites which survive atmospheric flight and are recovered, which are measured to be between $20-40$ MPa \cite{flynn2018physical}. Rarely, meteoroids act like monoliths and show no evidence of fragmentation \cite{borovivcka2008carancas}, possibly due to a lack of internal cracking which is commonly invoked as the mechanism which causes fragmentation at lower strengths \cite{borovivcka2005physical}. As we model fragmentation directly and use both the light curve and the dynamics as a constraint, the derived values should be accurate to within $\pm25\%$, which is the short-term variation in the atmosphere mass density that is not captured by current atmosphere models \cite{vida2021high}.

\subsection*{Computing PE} \label{subsec:pe}

The PE criterion was derived as an empirical tool to help easily differentiate between different material types \cite{ceplecha1976fireball} without the complexity of full numerical ablation modelling. It was based on well understood relationships between physical properties of meteoroids and their observed behaviour as they enter the atmosphere.

It is defined as:

\begin{equation}
    PE = \log \rho_E - 0.42 \log m_0 + 1.49 \log v_0 - 1.29 \log \cos Z_C
\end{equation}

\noindent where $\rho_E$ is the atmosphere mass density at the end height of the fireball in $\mathrm{g~cm^{-3}}$, $m_0$ is the initial mass in grams, $v_0$ the initial velocity in $\mathrm{km~s^{-1}}$, and $Z_C$ is the zenith angle.

Note that for the PE criterion to be properly computed, the correct mass scale (as outlined in refs. \cite{ceplecha1976fireball} and \cite{ceplecha1988earth}) needs to be used. In general, initial meteoroid mass is computed by integrating the total observed light production, assuming a spectral distribution $P_{\mathrm{0m}}$ and a luminous efficiency $\tau$ \cite{vida2018modelling}:

\begin{equation} \label{eq:meteoroid_mass}
    m = \frac{2 P_{\mathrm{0m}}}{\tau v^2} \int_{0}^{t} 10^{-0.4 M(t)}
\end{equation}

\noindent where $M(t)$ is the observed magnitude. As it enters the atmosphere, a meteoroid only possesses kinetic energy. The luminous efficiency measures how much of that kinetic energy gets converted into light and is usually on the order of a few percent. 

PE was originally derived using the following luminous efficiency which should always be used for PE computation:

\begin{equation} \label{eq:c76}
    \tau = \left\{\begin{array}{ll}
        1.5 \times 10^{-2.75}, & \text{for } v_0 \leq 9.3 \mathrm{km~s^-{1}} \, \\
        1.5 \times 10^{-5.60 + 2.92 \log v_0}, & \text{for } 9.3 < v_0 \leq 12.5~\mathrm{km~s^{-1}} \, \\
        1.5 \times 10^{-3.24 + 0.77 \log v_0}, & \text{for } 12.5 < v_0 \leq 17.0~\mathrm{km~s^{-1}} \, \\
        1.5 \times 10^{-2.50 + 0.17 \log v_0}, & \text{for } 17.0 < v_0 \leq 27.0~\mathrm{km~s^{-1}} \, \\
        1.5 \times 10^{-3.69 + 1.00 \log v_0}, & \text{otherwise} \, \\
        \end{array} \right.
\end{equation}

\noindent where $\tau$ is the dimensionless luminous efficiency as a fraction (not percentage), and $v_0$ the initial velocity in $\mathrm{km~s^{-1}}$.

These values of luminous efficiency are considered to be underestimated compared to contemporary models. For example, for the Alberta fireball with $v_0 = 62.1 \mathrm{km~s^{-1}}$ the luminous efficiency according to equation \ref{eq:c76} is $\tau = 1.9\%$, while in the modelling we used values between $10\%$ and $14\%$ (depending on the mass). To scale contemporary mass values to the appropriate values for PE computation, we simply scale the mass by the ratio of the historic (eq. \ref{eq:c76}) and modern \cite{borovivcka2020two} luminous efficiencies.

\subsection*{Strength of Cometary Material and Cometary Refractory Inclusions}

In-situ measurement of the nucleus of comet 67P by the Philae lander found a surface compressive strength of $1-3$~kPa \cite{biele2015landing}, a value consistent with in-atmosphere measurements of cometary meteoroid strength \cite{vida2021high}. The probe stopped bouncing when it hit an area of crushing strength $> 4$ MPa, possibly a processed, tightly packed ``sintered'' surface layer \cite{thomas1994crushing} created by space weathering. Despite having a high crushing strength, the surface layer was also found to have a high porosity of 30 to 65\%. The best fit to data was found by using a surface bulk density of $470 \pm 45~\mathrm{kg~m^{-3}}$ \cite{spohn2015thermal}.

% thus limiting the upper bulk density of the layer to $1600-2500~\mathrm{kg~m^{-3}}$ assuming an all-silicate surface. In reality the bulk density is likely even lower due to the presence of organic material \cite{strazzulla1991primordial}. 

To explain Philae lander measurements, modelling of sintering for the comet 67P has shown that a  hardened surface layer (compressive strength 8~MPa) several meters thick can be formed due to space weathering \cite{kossacki2015comet}. However, the model found that the tensile strength (what is traditionally measured during atmospheric entry of meteoroids \cite{spurny2012bunburra}) of this layer is an order of magnitude lower.

The possibility of meteorite delivery from the outer Solar System was discussed in detail by ref. \cite{gounelle2008meteorites}. Using a numerical model which included radioactive decay  of  short-lived nuclides and exothermic crystallization of amorphous water ice to crystalline ice as sources of heat, they explored if it was possible to sustain liquid water in comets for 1~kyr$-$1~Myr shortly after the formation of the Solar System, long enough for the hydrothermal alteration to produce CI chondritic material. They concluded that the transformation was possible, but discuss no possibility of forming ordinary chondritic material. We are unaware of any proposed physical process in the literature which can transform soft cometary material into material of similar bulk density and strength to ordinary chondrites.

% Static penetration resistances between 6.5 and 10~MPa can be expected if the surface was exclusively composed of sintered $\mathrm{CO_2}$ or $\mathrm{H_2O}$ granular ice at 220~K with porosities of 50 to 70\% \cite{komle2001impact}. 

Based on all of the foregoing considerations, macroscopic samples of cometary surface layers are expected to have smaller bulk densities than monolithic silicate material. Nevertheless, cometary material can have mm$-$sized inclusions of a stronger material such as calcium-aluminum-rich inclusions (CAIs), as found in the dust of comets Wild 2 \cite{joswiak2017refractory} and 67P \cite{paquette2016searching}. In addition, some mm-sized components of Taurid meteoroids (from 2P/Encke) were found to withstand pressures of up to 300 kPa \cite{matlovivc2017spectra, borovivcka2020physical}, and strong mm-sized inclusions have been found in Leonid meteoroids (from an HTC 55P/Tempel-Tuttle)\cite{borovivcka2000time}. 

Most recently, a survey \cite{vojavcek2019properties} of mm-sized meteoroids which analyzed their spectral and fragmentation properties, identified two iron meteoroids on HTC orbits (out of a total of 64 HTC meteors). The authors suggested these were ejected during the formation of the Solar Solar due to the dynamical instability caused by Jupiter's migration.

\subsection*{Astrometric Calibration} \label{sec:astrometric_calibration}

\begin{figure*}[!htbp]
		\includegraphics[width=0.75\textwidth]{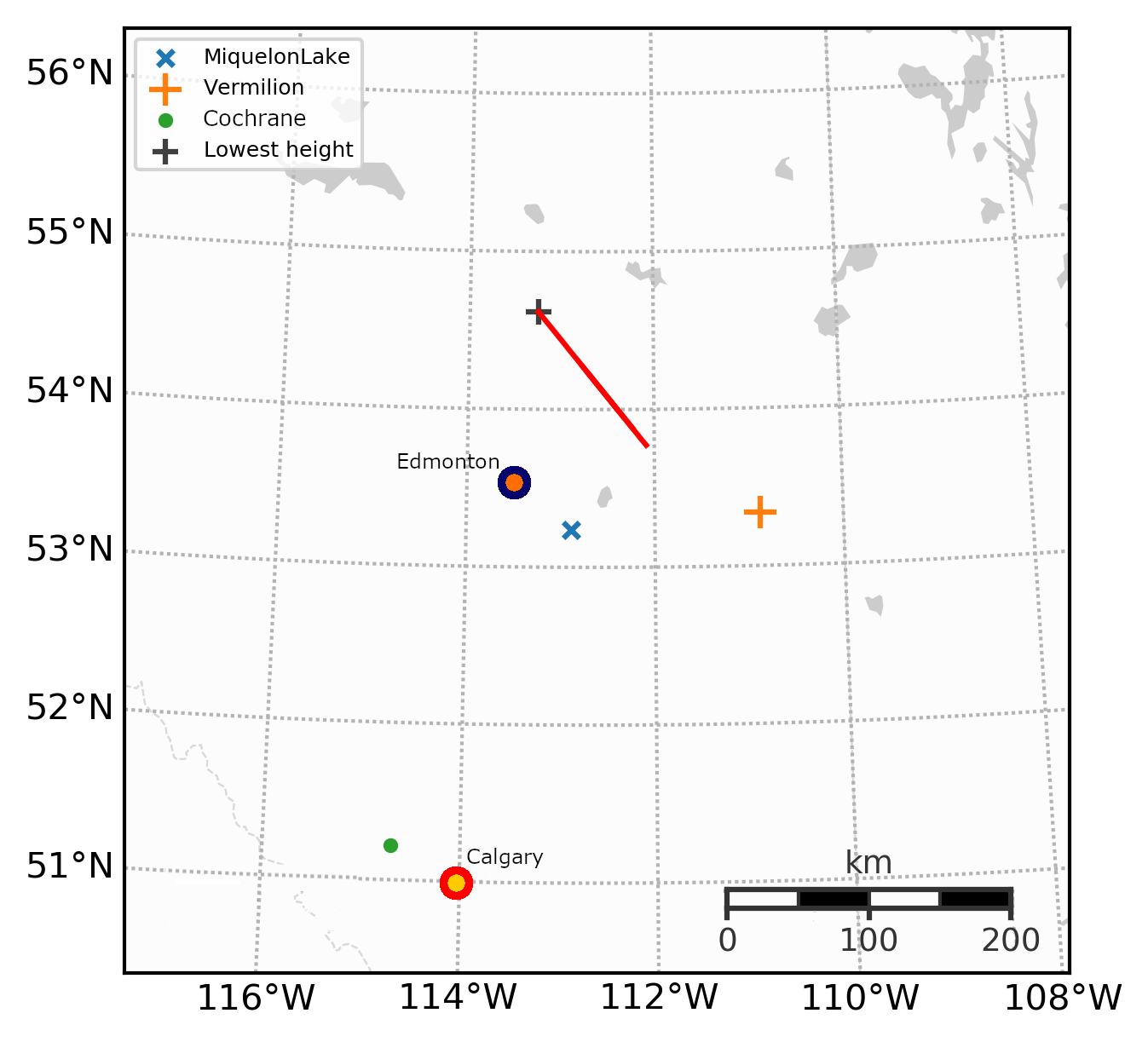}
	\caption{Map showing the location of the fireball trajectory (red line), cameras, and major population centers in Calgary and Edmonton. The GoPro camera was located in Calgary.}
	\label{fig:map}
\end{figure*}

The most critical measurement leading to the core result in this work (the unusually low end height for such a high velocity fireball) is directly derived from optical observations of the event. Thus, the quality of the astrometric calibration and measurements is of paramount importance. In this section, we present the calibration details for each of the three optical instruments used to derive the trajectory. These include two dedicated high resolution Global Fireball Observatory (GFO) fireball cameras (one at Miquelon Lake and the other near Vermilion, Alberta) and one security camera (located in Cochrane, Alberta, outside of Calgary). The camera locations are given in Table \ref{tab:camera_locations} and shown in relation to the fireball in Figure \ref{fig:map}.

\begin{table*}[!hb]
\caption{Geographical coordinates of optical cameras used in this work.}
\label{tab:camera_locations}
\begin{tabular}{lllll}
Name & Field of view & Latitude (+N) & Longitude (+E) & Elevation (mean sea level) \\
\hline\hline % inserts double horizontal lines
GFO Miquelon Lake & all-sky & 53.239340$^{\circ}$ & -112.889855$^{\circ}$ & 785~m \\
GFO Vermilion     & all-sky & 53.338975$^{\circ}$ & -110.884085$^{\circ}$ & 624~m \\
Cochrane          & $134^{\circ} \times 105^{\circ}$ & 51.224138$^{\circ}$ & -114.706921$^{\circ}$ & 1213~m \\
GoPro (photometry)  & $140^{\circ} \times 90^{\circ}$  & 50.910076$^{\circ}$ & -114.038749$^{\circ}$ & 1040~m
\end{tabular}
\end{table*}

\subsubsection*{Global Fireball Observatory Data}

The GFO all-sky cameras operated by the MORP2.0 project produce $7340 \times 4930$~px color images with an exposure time of 27~s and 14 bits of depth. Electronic liquid crystal shutters are toggled to encode the timing information into the image. The shutters produce 20 segments per second, and the segments are encoded as a de Brujin sequence of ones and zeros \cite{howie2017submillisecond} so that the absolute time of every segment can be derived to an accuracy of 1~ms. In combination with the Samyang 8~mm f/3.5 fish-eye lens, the images have a plate scale of $\mathrm{2~arcmin~px^{-1}}$. The data produced by these cameras are of similar quality and have a similar sensitivity limit as compared to other fireball networks.

The astrometric fit was performed with a radial distortion model \cite{vida2021global} using odd polynomial coefficients up to the 7$^\mathrm{th}$ order, asymmetry correction, and a fixed aspect ratio. Including the pointing direction (reference right ascension, declination, position angle, and the plate scale), the fit uses a total of 11 free parameters. 

\begin{figure*}[!htbp]
		\includegraphics[width=\textwidth]{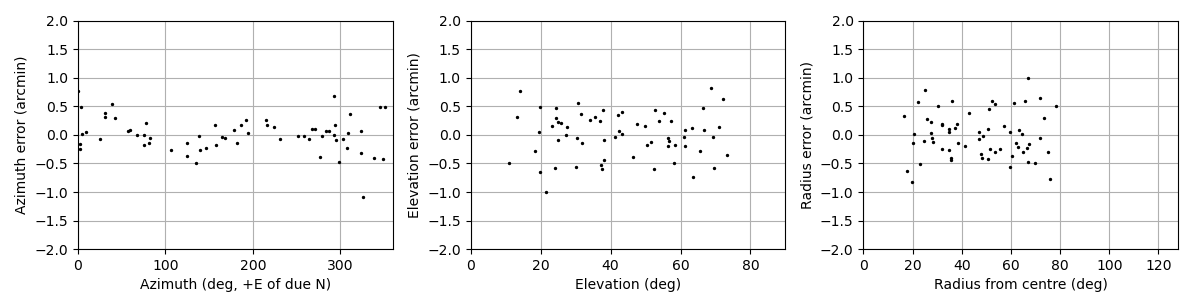}
	\caption{Astrometric calibration fit using a 7$^\mathrm{th}$ order polynomial (odd-terms only) radial distortion model for the Miquelon Lake camera. Forward mapping (image to sky) errors.}
	\label{fig:021_astrometry_fit}
\end{figure*}

Figure \ref{fig:021_astrometry_fit} shows the fit residuals for the Miquelon Lake camera. The mean angular forward mapping (image to sky) error was 0.49~arcmin, with a fit showing no trends in residuals with radius from the centre of the image but a slight systematic trend in the azimuth. We believe that the main cause of the trend is a higher order component of asymmetry in the optics not captured by the distortion model, as the point-spread function significantly varied across the field of view. Nevertheless, this offset is only on the order of 0.5~arcmin, i.e. 20~m at the range of the fireball from the station, and does not significantly influence the final result. The fireball covered azimuths from $30^{\circ}$ to $350^{\circ}$ (counter-clockwise) and elevations from $54^{\circ}$ to $15.7^{\circ}$, ranges well covered by available calibration stars.

\begin{figure*}[!htbp]
		\includegraphics[width=\textwidth]{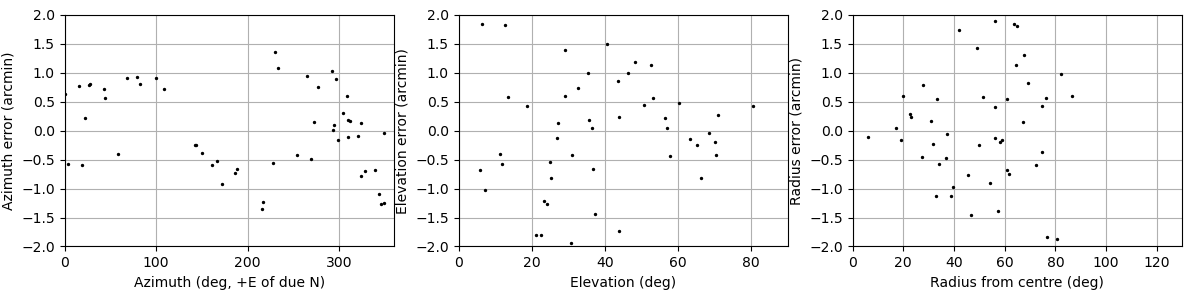}
	\caption{Astrometric calibration fit for the Vermillion camera. Forward mapping (image to sky) errors.}
	\label{fig:044_astrometry_fit}
\end{figure*}

The astrometric fit for the Vermilion GFO station was not as good, having root mean square fit residuals of 1.37~arcmin, as shown in Figure \ref{fig:044_astrometry_fit}. This decrease in accuracy was caused by a more non-Gaussian point-spread function than for the Miquelon Lake camera. However, the absolute accuracy remained high $-$ the corresponding linear error was only 65~m at the range of the fireball. The fireball had a nearly constant azimuth of $310^{\circ}$ and covered elevations from $43^{\circ}$ to $11^{\circ}$, ranges for which there were many stars in the calibration.

\subsection*{Security Camera Calibration}

Despite the good geometry and high accuracy of GFO measurements (convergence angle of $46.5^{\circ}$ and spatial trajectory fit residuals of $\sim30$~m), a two-station trajectory solution can suffer from systematic biases due to meteor-station geometry \cite{vida2020estimating, vida2020results}. As a further constraint, we included additional measurements from a Google Nest doorbell camera in Cochrane, Alberta, 50~km west of Calgary.

\begin{figure*}[!htbp]
		\includegraphics[width=\textwidth]{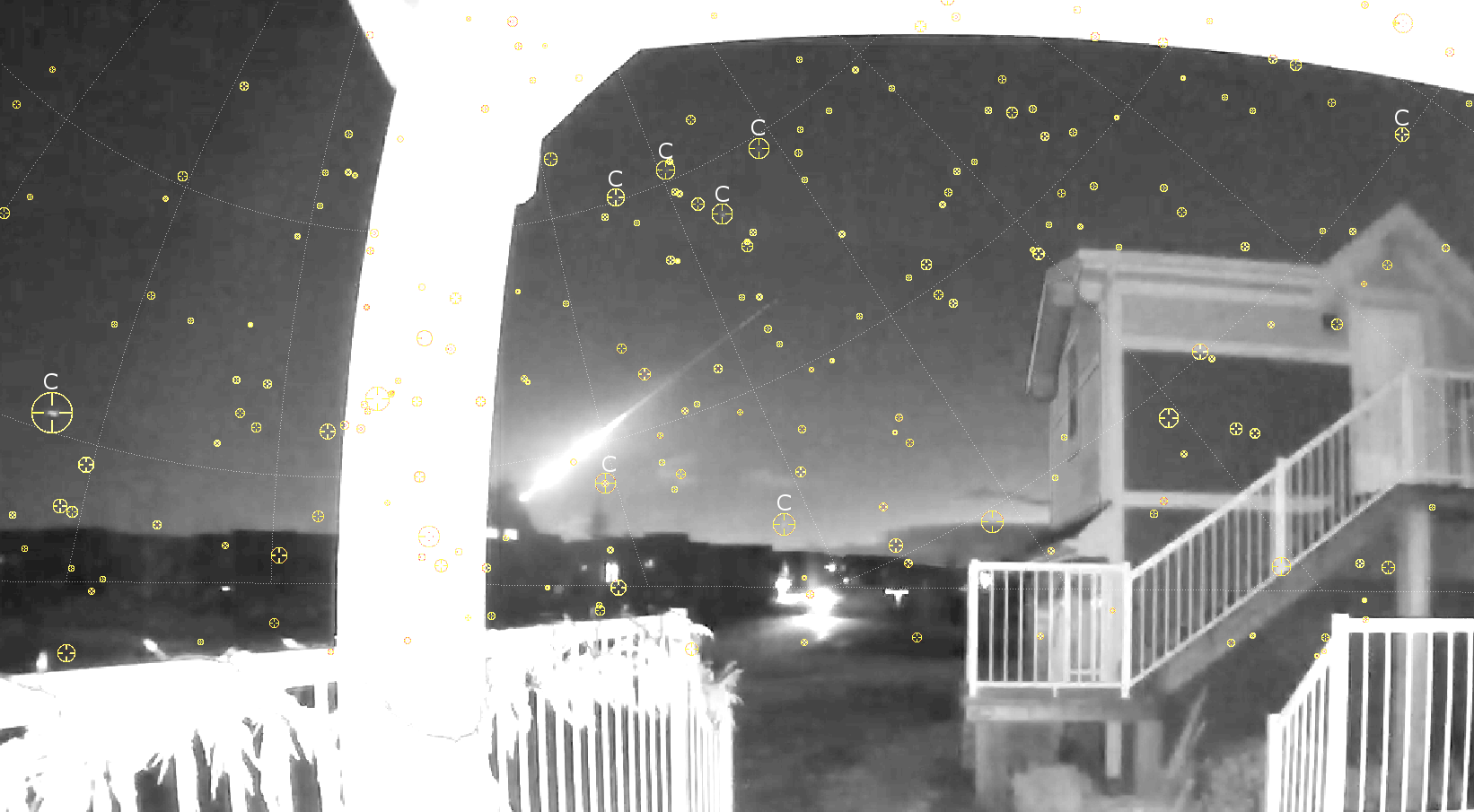}
	\caption{Composite of frames from the Cochrane security camera video showing the fireball and the calibration stars (marked with a white letter C), four of which were in Cassiopeia. An equatorial grid is laid over the video with catalog stars shown as yellow crosshairs. Credit: Airell DesLauriers.}
	\label{fig:airell_calib}
\end{figure*}

The radial distortion model with odd terms up to the fifth order was used for calibration \cite{vida2021global}. The model has a total of eight parameters and eight stars were used in the fit (Figure \ref{fig:airell_calib}). The average fit error was 4~arcmin. Only the first half of the fireball was used in the trajectory solution as the camera saturated and skipped frames during the brightest phase. The difference in the geocentric radiant with and without the security camera measurements was only $0.03^{\circ}$, and $0.14~\mathrm{km~s^{-1}}$ in geocentric speed, indicating that the GFO-only solution had no major systematic errors.

\subsection*{Photometric Calibration} \label{subsec:photometry}

To fully model the fragmentation behaviour of a meteoroid, it is necessary to have a well-calibrated light curve. The Alberta event was observed by both space-based GLM instruments and one fixed GoPro HERO5 action camera in Calgary ($\sim400$~km from the fireball). Note that the GFO cameras used for astrometry were partially saturated over the height range of interest and therefore not used in the photometric calibration. However, as the Nikon D810 used by the GFO have CMOS (and not CCD) chips for imaging, the astrometric position picks are reliable even in saturation \cite{vida2021global}.

The GoPro camera had a low sensitivity and in conjunction with the large range to the fireball it was able to capture the whole fireball without saturating despite only having 8 bits of depth. As the camera did not observe any stars, the absolute calibration was done indirectly using seven distant streetlights visible in the video (similar method was previously successfully applied in refs. \cite{spurny2010analysis, brown2019hamburg}). A separate digital single-lens reflex (DSLR) photograph of stars and the streetlights was taken by placing the DSLR camera next to the GoPro camera $-$ the apparent magnitudes of the streetlights were measured on the DSLR photo and used as a basis for the GoPro calibration. The mean photometric error was $\pm 0.27$~mag and the vignetting coefficient was estimated to be 0.001~$\mathrm{rad~px^{-1}}$ (see ref. \cite{vida2021global}).

An attempt was made to measure the photometry from scattered light on the Cochrane security camera video (method of ref. \cite{spurny2010analysis}), but the camera had a wide dynamic range (WDR) feature. This produces image levels that are not linear responses to light and was thus not able to be used. Such image enhancement features may prevent using modern security cameras for scattered light fireball photometry in the future.

\begin{figure*}[!htbp]
		\includegraphics[width=0.75\textwidth]{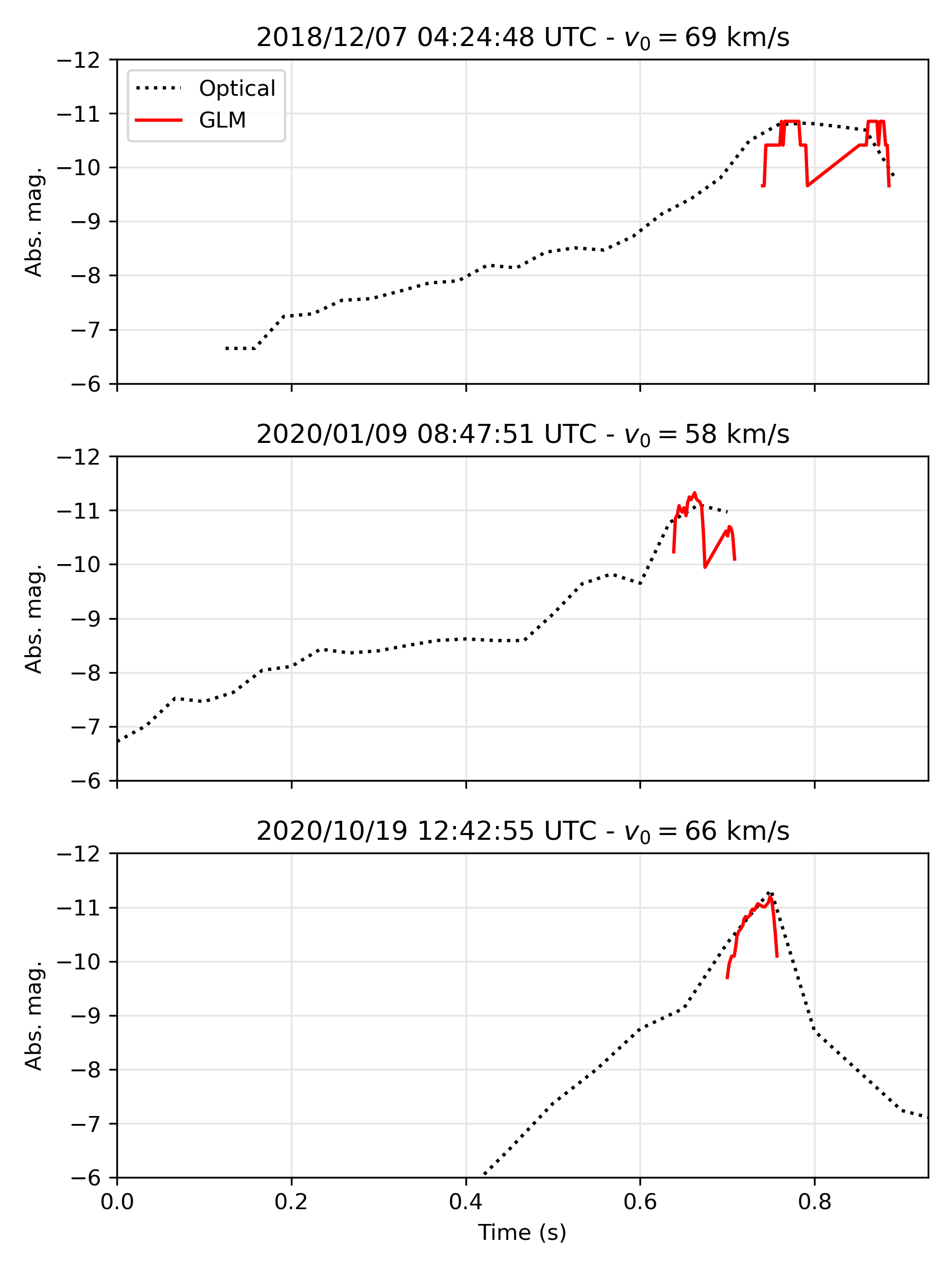}
	\caption{Comparison between optical light curves and GLM-derived light curves of calibration fireballs.}
	\label{fig:glm_lc_comparison}
\end{figure*}

Converting the energy observed by the GLM into magnitudes is challenging due to its narrow 1.1~nm pass band around 777.4~nm, making it necessary to assume a spectral energy distribution to compute a bolometric magnitude. For slower meteoroids such as meteorite-dropping fireballs, it is possible to assume a blackbody spectrum and derive a conversion \cite{jenniskens2018detection}, but at high speeds elemental and atmospheric lines are more pronounced making the blackbody assumption invalid. Furthermore, the intensity of the oxygen triplet line that the GLM is observing was found to significantly increase with meteoroid speed \cite{segon2018meteors}. For these reasons, we performed a manual calibration between the GLM group energy and magnitude using three fast and bright (around $-11$~mag) fireballs observed by the NASA Meteoroid Environment Office \cite{cooke2011status} and All-sky Meteor Orbit System (AMOS) all-sky cameras \cite{toth2019amos}. Among many fireballs observed by these systems, only two fireballs observed with NASA systems had GLM light curves and were observed sufficiently far away not to saturate the cameras. Some saturated frames in the AMOS recording (2020/10/19 12:42:55 UTC) were corrected using a calibration curve for saturated pixels, based on calibrated measurements of bright planets and Moon in different phases. The in-atmosphere speeds of the fireballs were 58, 66, and $69~\mathrm{km~s^{-1}}$, comparable to the Alberta fireball. We used the classical equation to compute the magnitude:

\begin{equation} \label{eq:glm_calib}
    M = -2.5 \log E_G + p_0
\end{equation}

\noindent where $E_G$ is the GLM group energy in femtojoules and the $p_0$ is the photometric offset in magnitudes. For all three fireballs, the GLM light curves matched best for $p_0 = -9.2$, with an error of $\pm 0.1$~mag. The comparison between the optical and GLM light curves is shown in Figure \ref{fig:glm_lc_comparison}.

\subsection*{Trajectory Details}

\begin{figure*}[!htbp]
		\includegraphics[width=0.5\textwidth]{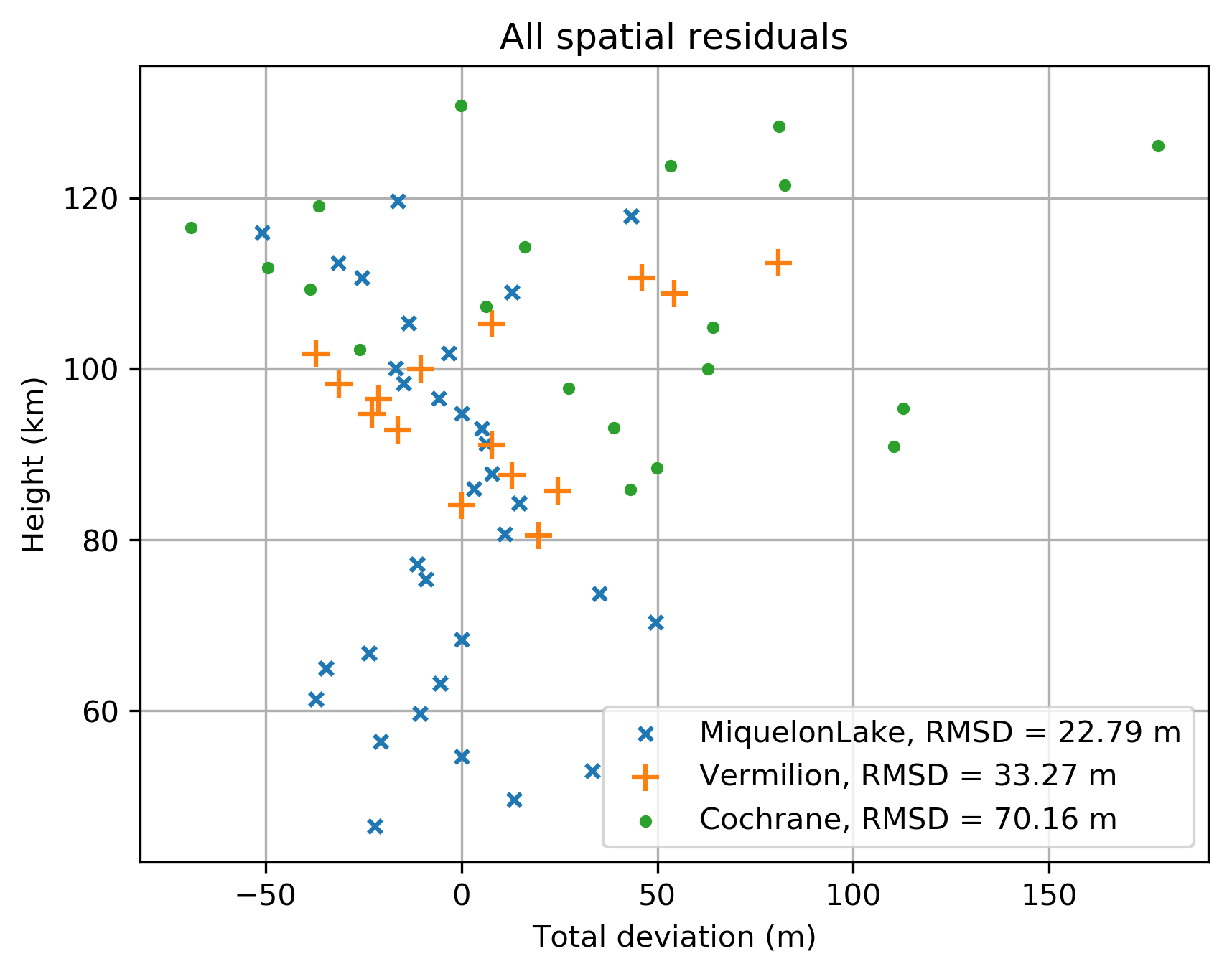}\hfill
		\includegraphics[width=0.5\textwidth]{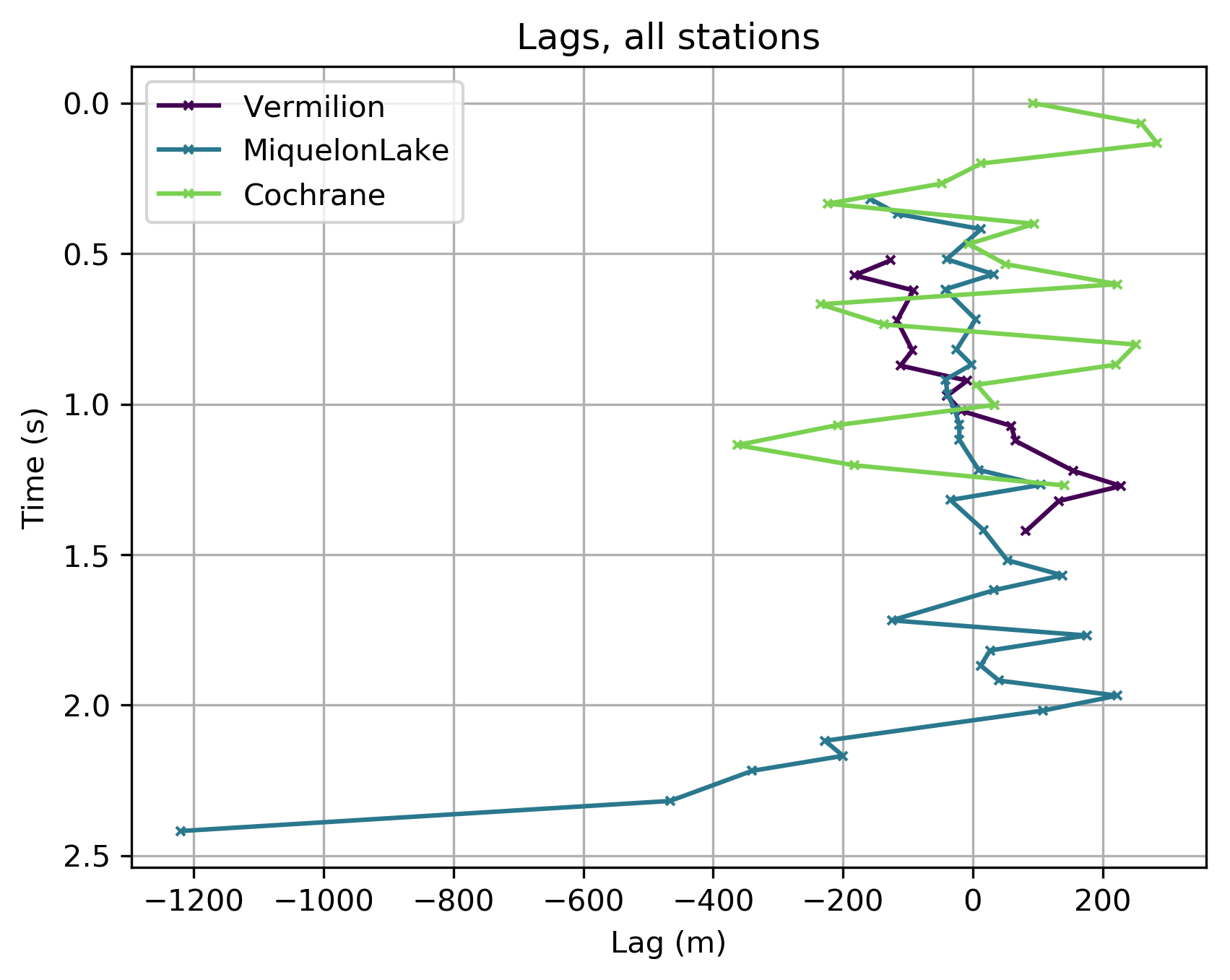}
	\caption{Left: Spatial trajectory fit residuals versus height. Right: The observed lag (``the distance that the meteoroid falls behind an object with a constant velocity that is equal to the initial meteoroid velocity''\cite{subasinghe2017luminous}.)}
	\label{fig:fit_errors_lag}
\end{figure*}

Table \ref{tab:trajectory} details the parameters of the begin and end point of the fireball. Figure \ref{fig:fit_errors_lag} shows the trajectory fit residuals and the observed deceleration. The trajectory fit is tight and within the expected astrometric accuracy. The fireball did not show much deceleration before a height of 60~km. The initial velocity was computed as the average velocity above the height of 70~km.

\begin{table*}
\caption{Fireball trajectory parameters.}
\label{tab:trajectory}
\begin{tabular}{lrrrl}
& Latitude (+N) & Longitude (+E) & Height (km, WGS84) & Time (UTC)\\
\hline\hline % inserts double horizontal lines
Begin point & $53.7732^{\circ} \pm 0.092~\mathrm{km}$ & $-112.0861^{\circ} \pm 0.036~\mathrm{km}$ & $130.819 \pm 0.082$ & 2021-02-22 13:23:17.683\\
End point   & $54.6182^{\circ} \pm  0.213~\mathrm{km}$ & $-113.2543^{\circ}  \pm 0.040~\mathrm{km}$ & $46.498 \pm 0.066$ & 2021-02-22 13:23:20.101\\
\end{tabular}
\end{table*}

The reference time for the trajectory is 2021-02-22 13:23:17.683 UTC (Julian date 2459268.057843548711). The state vector in the Earth-centered inertial (ECI) coordinates in the epoch of date is given in Table \ref{tab:state_vector}, the state vector covariance matrix is given in Table \ref{tab:state_vector_cov}, and the orbital covariance matrix is given in Table \ref{tab:orbit_cov}.

\begin{table*}
\centering
\caption{ECI trajectory state vector.}
\label{tab:state_vector}
\begin{tabular}{rrlrl}
    $X =$ & -1844963 & $\pm$ & 44 & $\mathrm{m}$ \\
    $Y =$ & -3384728 & $\pm$ & 193 & $\mathrm{m}$ \\
    $Z =$ &  5227379 & $\pm$ & 153 & $\mathrm{m}$ \\
    $V_X =$ &    2161 & $\pm$ & 41 & $\mathrm{m/s}$ \\
    $V_Y =$ &  -61705 & $\pm$ & 15 & $\mathrm{m/s}$ \\
    $V_Z =$ &    5244 & $\pm$ & 155 & $\mathrm{m/s}$
\end{tabular}
\end{table*}

\begin{table*}
\centering
\caption{ECI state vector covariance matrix. }
\label{tab:state_vector_cov}
\begin{tabular}{lrrrrrr}
      &            $X$ (m)       &         $Y$ (m)          &            $Z$ (m) &         $V_X$ (m/s)      &         $V_Y$ (m/s)   &          $V_Z$ (m/s)  \\
$X$   & $+1.778\mathrm{e}{+03}$ & $+4.430\mathrm{e}{+02}$ & $+4.577\mathrm{e}{+03}$ & $+1.499\mathrm{e}{+03}$ & $+8.165\mathrm{e}{+02}$ & $+4.720\mathrm{e}{+03}$ \\
$Y$   & $+4.430\mathrm{e}{+02}$ & $+3.851\mathrm{e}{+04}$ & $-7.591\mathrm{e}{+02}$ & $+1.377\mathrm{e}{+03}$ & $+4.232\mathrm{e}{+03}$ & $+2.417\mathrm{e}{+03}$ \\
$Z$   & $+4.577\mathrm{e}{+03}$ & $-7.591\mathrm{e}{+02}$ & $+2.168\mathrm{e}{+04}$ & $+4.334\mathrm{e}{+03}$ & $+2.671\mathrm{e}{+03}$ & $+2.039\mathrm{e}{+04}$ \\
$V_X$ & $+1.499\mathrm{e}{+03}$ & $+1.377\mathrm{e}{+03}$ & $+4.334\mathrm{e}{+03}$ & $+1.491\mathrm{e}{+03}$ & $+9.583\mathrm{e}{+02}$ & $+4.765\mathrm{e}{+03}$ \\
$V_Y$ & $+8.165\mathrm{e}{+02}$ & $+4.232\mathrm{e}{+03}$ & $+2.671\mathrm{e}{+03}$ & $+9.583\mathrm{e}{+02}$ & $+1.448\mathrm{e}{+03}$ & $+3.880\mathrm{e}{+03}$ \\
$V_Z$ & $+4.720\mathrm{e}{+03}$ & $+2.417\mathrm{e}{+03}$ & $+2.039\mathrm{e}{+04}$ & $+4.765\mathrm{e}{+03}$ & $+3.880\mathrm{e}{+03}$ & $+2.163\mathrm{e}{+04}$ \\
\end{tabular}
\end{table*}

\begin{table*}
\centering
\caption{Orbital covariance matrix. $T_p$ is the Julian date of last perihelion (nominal $T_p$ = 2459230.729594).}
\label{tab:orbit_cov}
\begin{tabular}{lrrrrrr}
         &       $e$                &    $q$ (au)              &     $T_p$ (day)          &      $\Omega$ (deg)      &       $\omega$ (deg)     &     $i$ (deg)            \\
$e$      &  $+8.321\mathrm{e}{-06}$ &  $+4.075\mathrm{e}{-06}$ &  $+2.410\mathrm{e}{-04}$ &  $+1.762\mathrm{e}{-08}$ &  $+6.764\mathrm{e}{-04}$ &  $-5.763\mathrm{e}{-04}$ \\
$q$      &  $+4.075\mathrm{e}{-06}$ &  $+2.367\mathrm{e}{-06}$ &  $+1.240\mathrm{e}{-04}$ &  $+1.009\mathrm{e}{-08}$ &  $+3.757\mathrm{e}{-04}$ &  $-3.302\mathrm{e}{-04}$ \\
$T_p$    &  $+2.410\mathrm{e}{-04}$ &  $+1.240\mathrm{e}{-04}$ &  $+7.079\mathrm{e}{-03}$ &  $+5.341\mathrm{e}{-07}$ &  $+2.031\mathrm{e}{-02}$ &  $-1.746\mathrm{e}{-02}$ \\
$\Omega$ &  $+1.762\mathrm{e}{-08}$ &  $+1.009\mathrm{e}{-08}$ &  $+5.341\mathrm{e}{-07}$ &  $+5.399\mathrm{e}{-11}$ &  $+1.607\mathrm{e}{-06}$ &  $-1.767\mathrm{e}{-06}$ \\
$\omega$ &  $+6.764\mathrm{e}{-04}$ &  $+3.757\mathrm{e}{-04}$ &  $+2.031\mathrm{e}{-02}$ &  $+1.607\mathrm{e}{-06}$ &  $+6.030\mathrm{e}{-02}$ &  $-5.258\mathrm{e}{-02}$ \\
$i$      &  $-5.763\mathrm{e}{-04}$ &  $-3.302\mathrm{e}{-04}$ &  $-1.746\mathrm{e}{-02}$ &  $-1.767\mathrm{e}{-06}$ &  $-5.258\mathrm{e}{-02}$ &  $+5.786\mathrm{e}{-02}$
\end{tabular}
\end{table*}

\subsection*{Orbital Integration}

\begin{figure*}[!htbp]
		\includegraphics[width=\textwidth]{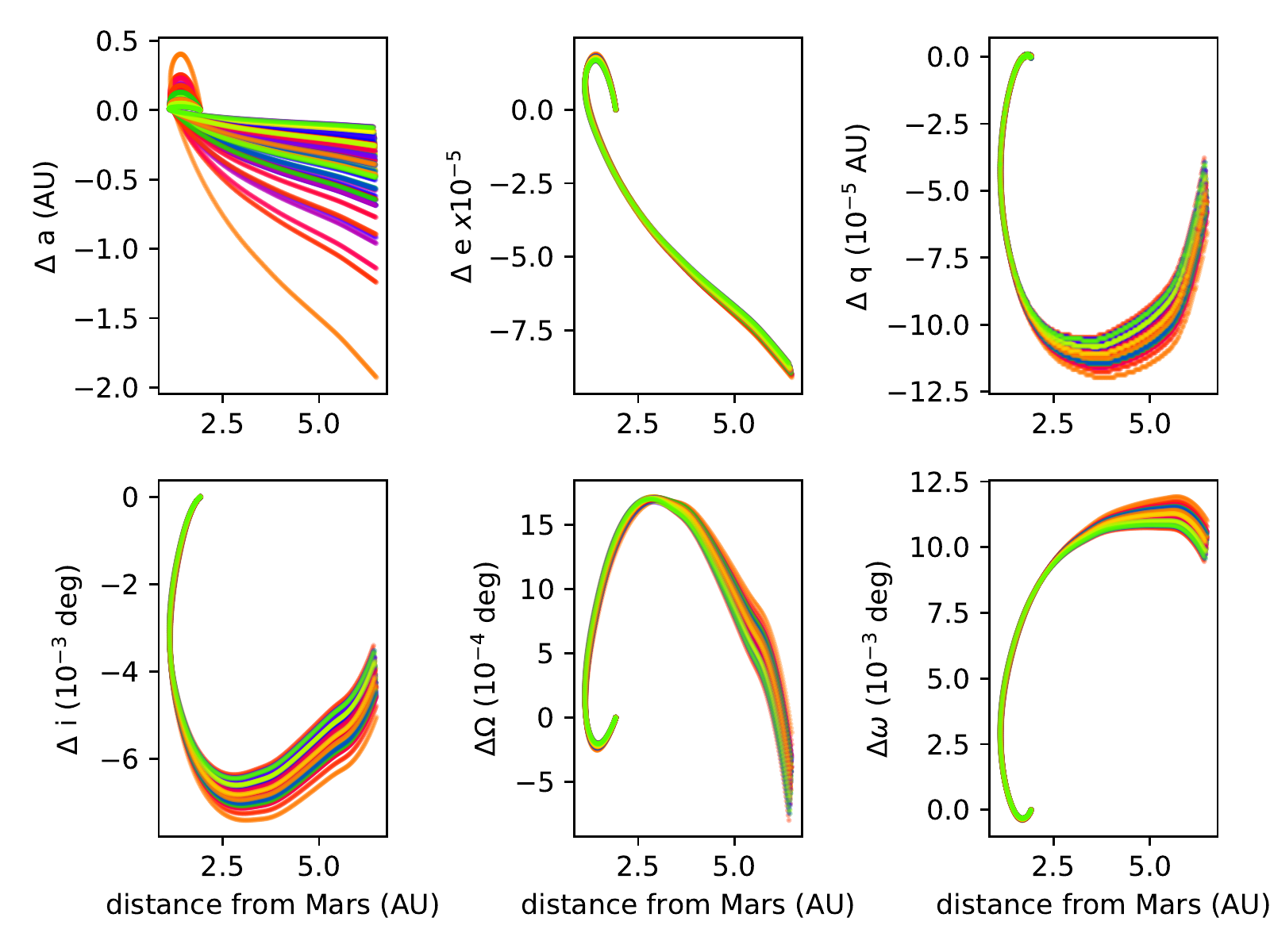}
	\caption{Change in orbital elements over time, between -60 and -365 days before impact. Each clone is color-coded individually and represents one sample within the orbital covarience matrix. Time is not shown on any axis, but the clones that start at t - 60 days are clustered at zero and spread out as we go further back in time, as the distance from Mars decreases and then increases again.}
	\label{fig:mars_encounter}
\end{figure*}

To investigate the influence of planetary interactions with the meteoroid's orbit, we backtracked 100 clones within the measured uncertainty. The RADAU \cite{everhart1985} 15th-order integrator was used with an error tolerance of $10^{-12}$ and an external time step of 1 day. Because of its high inclination, the only appreciable approaches to the planets are at its other (ascending) node, which is near Mars' orbit. However, no clone passed closer than 1.1 AU from this planet. Figure \ref{fig:mars_encounter} 
shows the variation in the heliocentric elements as a function of distance from Mars. The simulations start 60 days prior to impact and run for 365 days further back. A slight jump in each of the elements can be seen at the minimum distance from Mars, but it is small: it will not affect a potential parent body search and does not affect the proposed origin of the object.

Note that the orbital elements do not become completely constant even well after the Mars encounter. This is because they are heliocentric elements, and for such a large semi-major axis orbit, Jupiter's tug both on the Sun and the object create ongoing small changes. Even if the orbital elements were considered in the barycentric frame, the effect of ongoing planetary perturbations on such loosely bound orbits means that the orbital elements will have trends over time regardless of the reference frame.

\begin{figure*}[!htbp]
		\includegraphics[width=\textwidth]{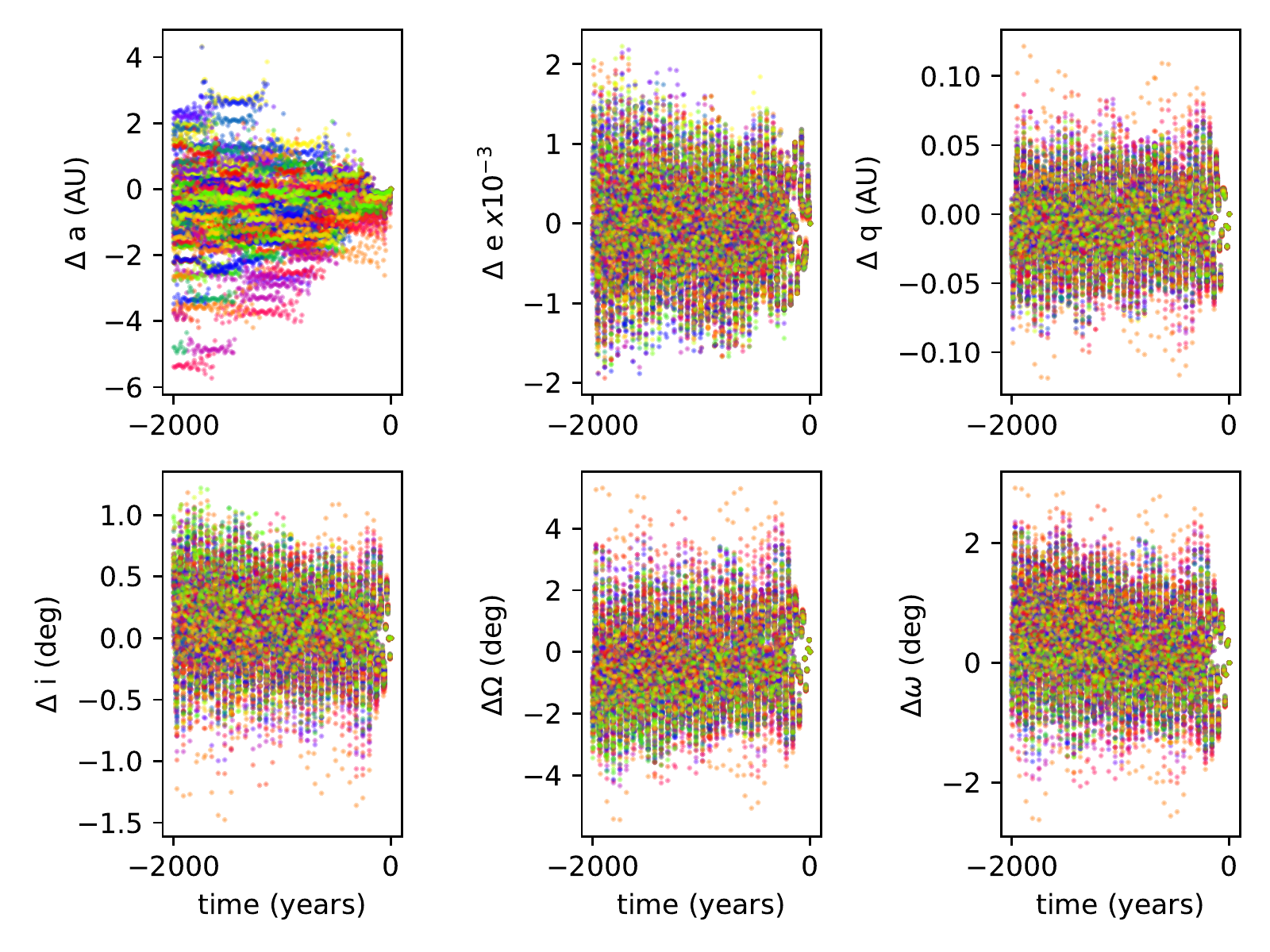}
	\caption{Change in orbital elements over time during a 2000 year backwards integration with all planets included. Each clone is color-coded individually. Planetary perturbations produce small nearly-stochastic changes in the orbital elements.}
	\label{fig:orbital_integration}
\end{figure*}

An extended integration backwards over 2000 years, corresponding to several orbits of the clones, reveals that planetary perturbations have had only a small effect on the orbital elements of the meteoroid over this time span (Figure \ref{fig:orbital_integration}).

\subsection*{Ablation Modelling Results} \label{subsec:model_results}

\begin{figure*}[!htbp]
		\includegraphics[width=0.75\textwidth]{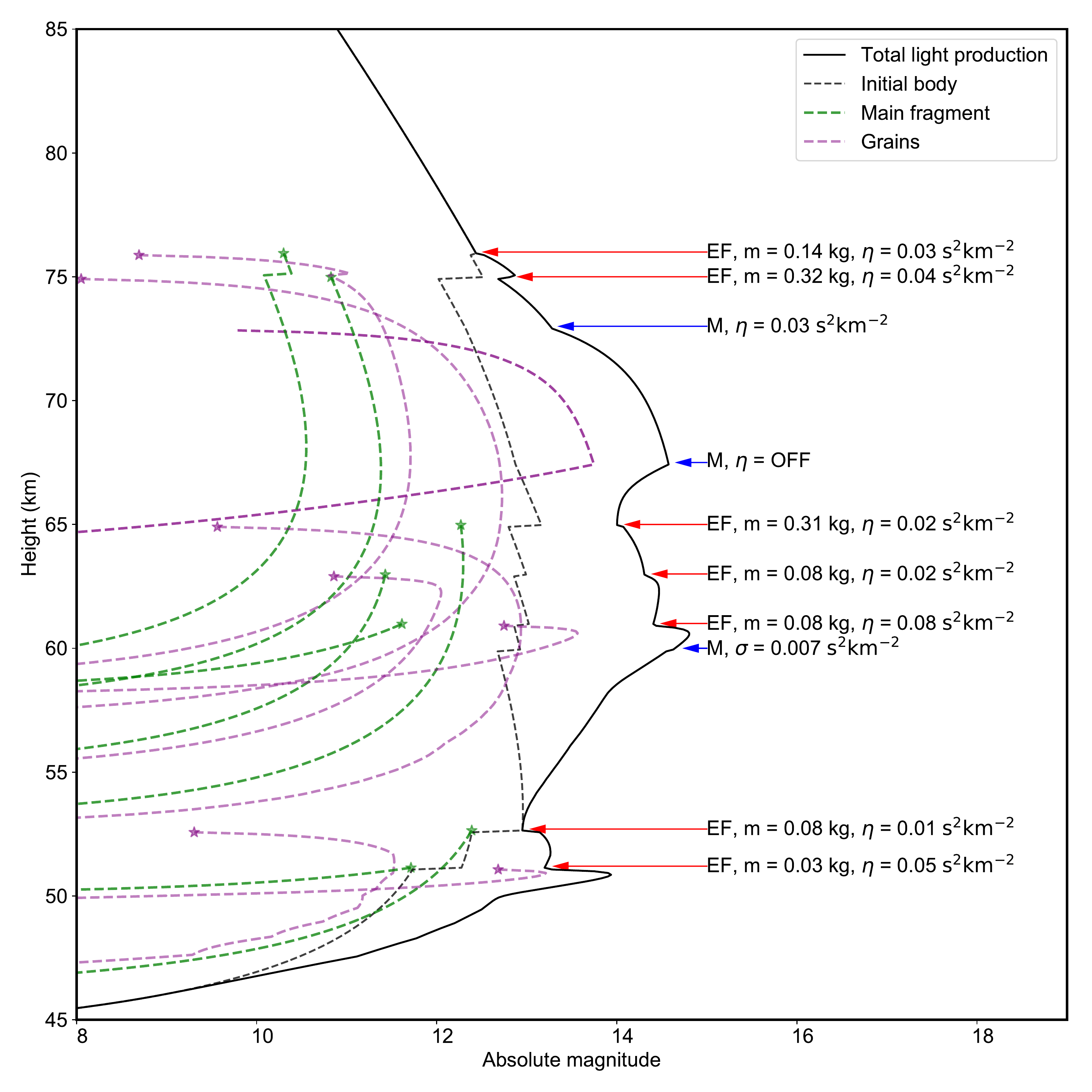}
	\caption{Details of the modelled individual fragmentations of the meteoroid marked on the simulated light curve. The solid black line is the total light production, the dashed black line is the magnitude of the main body from which fragments are released, green dashed lines are the magnitudes of the eroding fragments, and purple lines are the magnitudes of the grains ejected either from the main body or the eroding fragments. Arrows indicate where the fragmentations occurred with which parameters, and stars indicate the beginning of individual fragment/grain light curves.}
	\label{fig:frag_details}
\end{figure*}

\begin{figure*}[!htbp]
	    \includegraphics[width=0.75\textwidth]{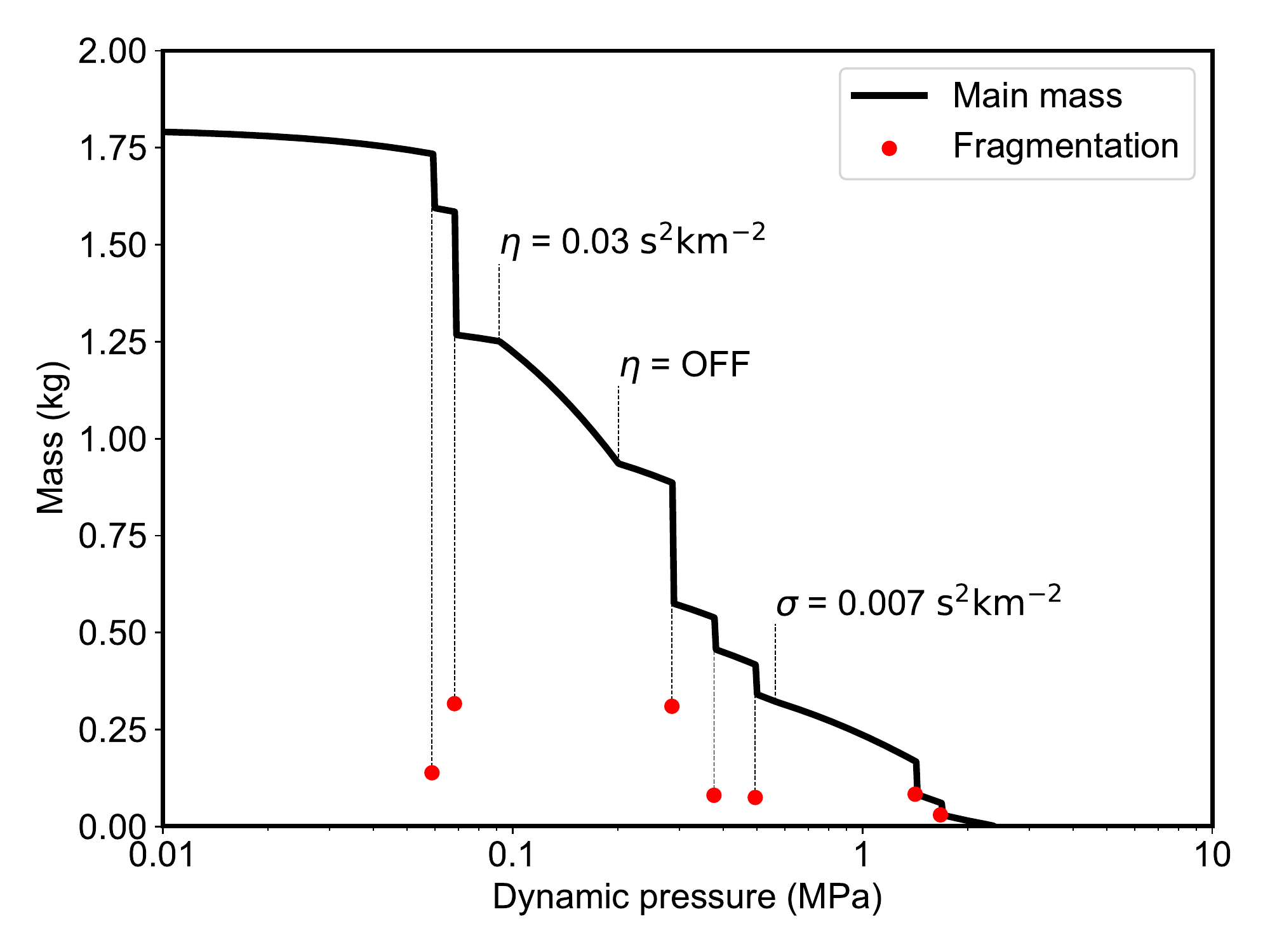}
	\caption{Modelled mass loss as a function of increasing dynamic pressure. Model fragmentation points and masses of major fragments are marked with red circles. $\eta$ marks the change in the erosion coefficient, and $\sigma$ the change in the ablation coefficient of the main body.}
	\label{fig:dyn_press_vs_mass_loss}
\end{figure*}

We model the fragmentation behaviour of the fireball in two ways. (1) by direct erosion of the main body (EM), and (2) by ejecting larger fragments which erode independently (EF). Figure \ref{fig:frag_details} 
shows the details of the fragmentation on the simulated light curve which are also listed in Table \ref{tab:fragmentation}, and Figure \ref{fig:dyn_press_vs_mass_loss} 
shows the mass loss with increasing dynamic pressure. The modelled physical properties of the fireball are given in Table \ref{tab:model_phys}. We only provide a single solution with no error estimate. The model is fit manually and a still unresolved question in the field is how to provide meaningful model uncertainties. The model is highly non-linear and defining a robust cost function has also not yet been addressed. Previous attempts to automate the model fits ignored either the dynamics or the photometric measurements and failed to model the fragmentation directly \cite{sansom2015novel, sansom20193d, tarano2019inference}. In our approach, we use the dynamics as a hard constraint to accurately classify the material type and include additional fragmentation details to explain the light curve. This established method is often used to accurately model meteorite-dropping fireballs and accurately predict the masses of meteorites on the ground \cite{borovivcka2013kovsice, borovivcka2015instrumentally}.

\begin{table*}
\caption{Modelled fragmentation behaviour. The total mass of all ejected fragments is 1.04~kg. The fragment mass percentage in the table is reference to the mass of the main fragment at the moment of ejection. The mass distribution index for all grains was $s = 2.0$. The values of the dynamic pressure are computed using the drag coefficient $\Gamma$ used in the modelling for the appropriate height, as described in Supplementary Table 7.
%\ref{tab:model_phys}
}
\label{tab:fragmentation}
\begin{tabular}{lllllrrlllll}
Time$\mathrm{^{a}}$  & Height & Velocity               &  Dyn pres & Main $m$ & Fragment  & $m$ & $m$     & Erosion coeff            & Grain $m$  \\
(s)   & (km)   & ($\mathrm{km~s^{-1}}$) &  (MPa)    & (kg)       &       & (\%) & (kg)   & ($\mathrm{kg~MJ^{-1}}$) & range (kg) \\
\hline\hline % inserts double horizontal lines
1.49 & 76.0   & 62.03                  &  0.09     &    1.73 &    EF &    8 & 0.139 & 0.030                    &  $10^{-7}$ - $10^{-6}$ \\
1.52 & 75.0   & 62.02                  &  0.10     &    1.59 &    EF &   20 & 0.317 & 0.035                    &  $10^{-6}$ - $10^{-4}$ \\
1.57 & 73.0   & 62.00                  &  0.10     &    1.25 &    EM &    - &     - & 0.030                    &  $10^{-6}$ - $10^{-5}$ \\
1.72 & 67.5   & 61.88                  &  0.21     &    0.93 &    EM &    - &     - & OFF                      &          -             \\
1.79 & 65.0   & 61.78                  &  0.30     &    0.87 &    EF &   35 & 0.310 & 0.020                    &  $10^{-5}$ - $10^{-3}$ \\
1.84 & 63.0   & 61.66                  &  0.38     &    0.54 &    EF &   15 & 0.081 & 0.020                    &  $10^{-7}$ - $10^{-6}$ \\
1.90 & 61.0   & 61.49                  &  0.48     &    0.42 &    EF &   18 & 0.075 & 0.080                    &  $10^{-7}$ - $10^{-6}$ \\
2.12 & 52.7   & 59.84                  &  1.28     &    0.17 &    EF &   50 & 0.084 & 0.005                    &  $10^{-4}$ - $10^{-3}$ \\
2.17 & 51.2   & 59.07                  &  1.52     &    0.06 &    EF &   50 & 0.030 & 0.050                    &  $10^{-6}$ - $10^{-5}$ \\
\hline
\multicolumn{9}{l}{$\mathrm{^{a}}$ Seconds after 13:23:17.683 UTC.} \\
\multicolumn{9}{l}{EM = Main fragment erosion; EF = New eroding fragment.} \\
\end{tabular}
\end{table*}

\begin{table*}
\caption{Modelled physical properties of the fireball.}
\label{tab:model_phys}
\begin{tabular}{llrl}
Description & & Value\\
\hline\hline % inserts double horizontal lines
Initial mass (kg)                              & $m_0$    & 1.8  \\
Initial speed at 180~km ($\mathrm{km~s^{-1}}$) & $v_0$    & 62.10 \\
Zenith angle                                   & $Z_c$    & $54.885^{\circ}$ \\
Bulk density ($\mathrm{kg~m^{-3}}$)            & $\rho$   & 3300  \\
Grain density ($\mathrm{kg~m^{-3}}$)           & $\rho$   & 3500  \\
Ablation coefficient ($\mathrm{kg~MJ^{-1}}$)   & $\sigma$ & 0.009  \\
(below 60~km)                                  & $\sigma$ & 0.007  \\
Shape factor (sphere)                          & $A$      & 1.21   \\
Drag coefficient                               & $\Gamma$ & 0.8    \\
(below 75~km)                                  & $\Gamma$ & 0.6    \\
\end{tabular}
\end{table*}

Because this fireball is the first of its type to be ever modelled (large, high speed rocky meteoroid reaching low altitudes), there were several differences and uncertainties in chosen parameters as compared to low-velocity Type I objects:
\begin{enumerate}
    \item The luminous efficiency for Type I objects at high speeds is unknown, as this is the first object of this kind to be observed. We used the model of ref. \cite{borovivcka2020two} which suggest a value of $~\sim14\%$ for 1~kg objects and $~\sim10\%$ for grains. If the luminous efficiency is akin to low-speed meteorite dropping fireballs ($\sim5\%$), the initial mass is $\sim3\times$ larger (6~kg), but the identification of the meteoroid as a Type I fireball is unchanged. \\
    \item A higher intrinsic ablation coefficient of 0.009~$\mathrm{kg~MJ^{-1}}$ (and 0.007~$\mathrm{kg~MJ^{-1}}$ below 60~km) was used, as compared to 0.005~$\mathrm{kg~MJ^{-1}}$ for low-velocity deeply penetrating fireballs \cite{ceplecha1998meteor}. \\
    \item The model matches the light curve well even at the beginning of luminous flight at fainter magnitudes. This is not usually the case for slower meteorite-dropping fireballs as they undergo a period of preheating \cite{spurny2020vzvdar}, and the classical equations do not capture that complexity in those cases. \\
\end{enumerate}

\subsection*{Code availability}

The optical data were calibrated using the open source \texttt{SkyFit2} software available in the \texttt{RMS} library at URL: \url{https://github.com/CroatianMeteorNetwork/RMS}.

The \texttt{WesternMeteorPyLib} (\texttt{wmpl}) library was used to compute the trajectory and fit the meteoroid ablation model to the observations. It is available at URL: \url{https://github.com/wmpg/WesternMeteorPyLib/}.

\subsection*{Data availability}
The trajectory data are included in this article as Supplementary Data files. The raw images and supplementary are available on Zenodo: \url{https://doi.org/10.5281/zenodo.7225827}.

% *** END NOTES ***
\newpage
\section*{End Notes}
\subsection*{Acknowledgements}
We thank the two anonymous reviewers for their expert comments and valuable suggestions which greatly improved the quality of the this work. Also, we thank Mr. Robert Howell for first bringing this fireball to our attention and Mr. Airell DesLauriers for providing raw footage from his security camera in Cochrane, AB. 

Funding for this work was provided in part through NASA co-operative agreement 80NSSC21M0073 (DV, PGB), by the Natural Sciences and Engineering Research Council of Canada Discovery Grants program (Grants no. RGPIN-2016-04433 \& RGPIN-2018-05659, DV, PGB), the Canada Research Chairs program (PGB), the Slovak Research and Development Agency grant APVV-16-0148, and the Slovak Grant Agency for Science grant VEGA 1/0218/22 (PM, JT).

\subsection*{Author Contributions}
DV coordinated the effort, performed the analysis, implemented the software, and wrote the manuscript. PGB initially coordinated the effort and provided scientific insight. HARD computed an initial trajectory and provided the raw GFO data. PW made valuable scientific interpretations of the results and added a connection to recent comet discoveries, and performed the orbital integrations. DEM helped digitize the MORP data, systematically collected and organized all casual recordings of the fireball, provided the GLM observations, and identified fireballs for GLM calibration. PM and JT provided observations of a fireball jointly observed with the AMOS system and GLM. CDKH and PJAH provided the GFO data from the Miquelon Lake and Vermilion cameras and helped contact the local people who observed the fireball. EKS and MCT analyzed the global GFO data set to locate other fireballs of interest and provided data access. WJC provided initial coordination. DWH provided the GoPro video and took DSLR images for its photometric calibration.

\subsection*{Declaration of Competing Interests}
The authors declare no competing interests.

% % *** TABLES ***
% \newpage
% \section{Tables}

% % *** FIGURE LEGENDS ***
% \newpage
% \section{Figure Captions}

% \begin{itemize}
% 	\item[] \ref{fig:tj_vs_pe}: \textbf{\lipsum[20][1]} \textbf{(A)} \lipsum[12][1] \textbf{(B)} \lipsum[12][2] \textbf{(C)} \lipsum[12][3] \textbf{(D)} \lipsum[12][4] \textbf{(E)} \lipsum[12][5]
% 	\item[] blah: \textbf{\lipsum[20][2]} \lipsum[13][1-4]
% \end{itemize}

% *** REFERENCES ***
\newpage
\bibliography{bibliography}
\bibliographystyle{naturemag}

\end{document}